\documentclass[11pt,a4paper]{article}
\pdfoutput=1
\usepackage{dcolumn}
\usepackage{bm}
\usepackage[inline]{enumitem}
\usepackage{slashed}
\usepackage[caption=false]{subfig}
\usepackage{jheppub}

\DeclareMathOperator{\Li}{Li}
\DeclareMathOperator{\arctanh}{arctanh}

\def\gev{\, {\rm GeV}}
\def\mev{\, {\rm MeV}}
\def\kev{\, {\rm keV}}

\newcommand{\beq}{\begin{equation}}
\newcommand{\eeq}{\end{equation}}
\newcommand{\be}{\begin{equation}}
\newcommand{\ee}{\end{equation}}
\newcommand{\bea}{\begin{eqnarray}}
\newcommand{\eea}{\end{eqnarray}}

\title{Simplified Dark Matter Models with Charged Mediators: Prospects for Direct Detection}

\author[]{Pearl Sandick,}

\author[]{Kuver Sinha}

\author[]{and Fei Teng}

\emailAdd{sandick@physics.utah.edu}
\emailAdd{kuver.sinha@gmail.com}
\emailAdd{Fei.Teng@utah.edu}

\affiliation{
  Department of Physics and Astronomy, University of Utah, Salt Lake City, UT 84112, USA
}

\abstract{
We consider direct detection prospects for a class of simplified models of fermionic dark matter (DM) coupled to left and right-handed Standard Model fermions via two charged scalar mediators with arbitrary mixing angle $\alpha$. DM interactions with the nucleus are mediated by higher electromagnetic moments, which, for Majorana DM, is the anapole moment. After giving a full analytic calculation of the anapole moment, including its $\alpha$ dependence, and matching with limits in the literature, we compute the DM-nucleon scattering cross-section and show the LUX and future LZ constraints on the parameter space of these models. We then compare these results with constraints coming from $Fermi$-LAT continuum and line searches. Results in the supersymmetric limit of these simplified models are provided in all cases. We find that future direct detection experiments will be able to probe most of the parameter space of these models for $\mathcal{O}(100-200)$ GeV DM and lightest mediator mass $\lesssim \mathcal{O}(5\%)$ larger than the DM mass. The direct detection prospects dwindle for larger DM mass and larger mass gap between the DM and the lightest mediator mass, although appreciable regions are still probed for $\mathcal{O}(200)$ GeV DM and lightest mediator mass $\lesssim \mathcal{O}(20\%)$ larger than the DM mass. The direct detection bounds are also attenuated near certain ``blind spots" in the parameter space, where the anapole moment is severely suppressed due to cancellation of different terms. We carefully study these blind spots and the associated $Fermi$-LAT signals in these regions.
}

\begin{document}
\maketitle


\section{\label{sec:level1}Introduction}

The existence of dark matter (DM) in our universe has been established by various astrophysical and cosmological observations, notably galaxy rotation curves~\cite{Begeman:1991iy} and the cosmic microwave background~\cite{Ade:2013zuv}. The particle nature of DM is an area of intense study by both experimentalists and theorists, since it has the potential to illuminate several deep issues in the Standard Model (SM), such as the strong CP problem, neutrino masses, and the hierarchy problem. 

Weakly Interacting Massive Particles (WIMPs) are a well-motivated class of candidates that appear in theories beyond the SM, particularly those, like supersymmetry, that address the hierarchy problem. One finds that weak-scale couplings and masses give rise to a thermal relic density compatible with the measured DM density, lending further credence to these candidates. They have thus been extensively searched in colliders, as well as in indirect and direct detection experiments. A popular strategy for parametrizing WIMP searches is to work within ``simplified models", where one remains agnostic about the specific UV completion and allows for a wider coverage of theory parameter space. Aspects of simplified models with DM coupling to quarks and leptons have been investigated by many authors~\cite{Gershtein:2008bf, Godbole:2015gma, Bai:2010hh, Chang:2013oia, Bai:2013iqa, An:2013xka, DiFranzo:2013vra, Garny:2014waa, Papucci:2014iwa, Harris:2014hga, Chang:2014tea,Buckley:2014fba, Alves:2015dya,Baek:2015fma,Buckley:2015ctj}.

The purpose of this paper is to study simplified DM models with charged mediators, with a focus on direct detection constraints. For concreteness, we assume that the DM is a Majorana fermion. We consider two scalar mediators with arbitrary mixing angle $\alpha$, which couple the DM to left and right-handed SM leptons. Various aspects of this class of simplified models have been studied previously, such as constraints coming from SM electric and magnetic dipole moments, relic density, and indirect detection~\cite{Fukushima:2014yia,Kumar:2016cum}. Direct detection in the context of a similar class of models, but with colored mediators, has been studied by~\cite{Kelso:2014qja}. For uncolored charged mediators, DM interactions with the nucleus are mediated by higher electromagnetic moments. In the case of Majorana DM, the relevant one is the anapole moment. DM with anapole interactions have been studied in various contexts~\cite{Kopp:2014tsa,DelNobile:2014eta,Gresham:2013mua,Ho:2012br,Gao:2013vfa,Cabral-Rosetti:2015cxa}. 

Our main results include comparing direct and indirect detection constraints on the model, which are provided respectively by LUX~\cite{Akerib:2013tjd,Akerib:2015rjg,Akerib:2016vxi} and $Fermi$-LAT~\cite{Ackermann:2015lka,Ackermann:2015zua}. We provide a careful analytic calculation of the anapole moment and DM-nucleus scattering cross section. The constraints coming from LUX \footnote{As this paper was nearing completion, the LUX Collaboration put out new limits \cite{Akerib:2016vxi}. Results based on these limits, for which we give estimates, will be more stringent than the LUX 2014 bounds, but less so than the future LZ limits.}, and future projections from the LZ detector~\cite{Akerib:2015cja}, are then computed and shown for different choices of model parameters. On the indirect detection side, constraints both from gamma-ray line signals (when the DM and the mediator are sufficiently degenerate) and the continuum photon spectrum (for non-zero mixing of the mediators) are obtained. The calculations are done for representative DM masses 100 GeV and 200 GeV. Mixing angles $\alpha = 0, \pi/4,$ and $\pi/2$ are chosen when depicting constraints on other parameters. The mass eigenvalues of the two scalars are chosen to satisfy the existing collider limits. The special case of supersymmetry is highlighted throughout. 

Our results are highly sensitive to the level of degeneracy of the lightest mediator and the DM, which we parametrize as follows
\be
\mu \, = \, \frac{m^2_{\text{med.}}}{m^2_{\text{DM}}} \,\,.
\ee
Results for $\mu \, = \, 1.01, 1.10,$ and $1.44$ are displayed for each case.

We find that direct and indirect detection probe complementary regions of parameter space. For cases when the lightest mediator mass is most degenerate with the DM mass ($\mu = 1.01$), the direct detection limits from LUX constrain broad regions of parameter space and are comparable with current constraints from $Fermi$-LAT. As the lightest mediator becomes heavier and the mass gap with the DM increases, the magnitude of the anapole moment becomes smaller and consequently the direct detection limits become weaker. Indirect detection limits are largely indifferent to this mass separation, and start to dominate over LUX limits as the mediator becomes heavier for a given DM mass. 

We also note that future direct detection experiments will be very effective in probing the parameter space for these models. We take as an example the most optimistic projections from LZ, with $1$ background event in $1000$ days exposure of $5.6$ tonne fiducial mass, which is expected to lower the exclusion limit on the cross section by a factor of $7\times 10^{-4}$~\cite{Akerib:2015cja}.  With an improvement of three to four orders of magnitude in the effective DM-nucleus scattering cross section, the limits on the DM-mediator-SM fermion Yukawa coupling will get stronger by about one order of magnitude. Future indirect detection experiments like  GAMMA400~\cite{G400} and HERD~\cite{HERD}, on the other hand, expect an improvement by about a factor of several on the annihilation cross section, which only marginally improves the constraints on the Yukawa coupling. Thus, future direct detection constraints overwhelm indirect detection constraints.

It is important to point out that the results weaken considerably as either $\mu$ or the DM mass is increased. The supersymmetric limit of these simplified models is particularly interesting in this context. While future direct detection experiments will constrain the SUSY limit of the simplified model for 100 GeV DM with $\mu \leq 1.10$, choosing larger values of the DM mass (200 GeV) and mass gap ($\mu = 1.44$) leave the SUSY limit almost unconstrained. 

One aspect of these models that we examine carefully is the appearance of  certain ``blind spots'' in the parameter space, where the anapole moment is suppressed due to cancellation between various terms. This happens for certain choices of the mixing angle $\alpha$ and Yukawa couplings. Near these blind spots, the direct detection constraints become severely attenuated. We study the effectiveness of indirect detection in probing these regions.

Finally, we note that the focus of the present work is not to perform a full event-level analysis for the LUX 2016 results \cite{Akerib:2016vxi}, which we invite other groups to carry out in the future, but a careful calculation of the anapole-induced DM-nucleon scattering cross section in this particular model with charged mediators and the exploration of the interplay of constraints from direct and indirect DM searches. Thus, while we believe that the LUX 2016 and future LZ bounds that we have presented are broadly representative, the constraints would certainly be sharpened once a full event-level analysis is carried out with future datasets.

The rest of the paper is structured as follows. In Section \ref{sec:th}, we first discuss some general features of our simplified model. In Section \ref{sec:anapole} and Section \ref{sec:IB}, we discuss direct and indirect detection, respectively. In Section \ref{sec:result}, we present our results. We end with our conclusions, and present detailed calculations of the anapole moment and IB cross section in two Appendices.

\section{\label{sec:th}The Simplified Model}

We consider a Majorana DM candidate $\chi$ (with mass $m_{\chi}$) that couples only to an uncolored fermion $f$ (with mass $m_{f}$) and a pair of charged scalars $\widetilde{f}_{L,R}$. The interaction is described by the Lagrangian
\begin{equation}
\label{eq:Lint}
  \mathcal{L}_{\text{int}}=\lambda_{L}\widetilde{f}_{L}^{\ast}\overline{\chi}P_{L}f+\lambda_{R}\widetilde{f}_{R}^{\ast}\overline{\chi}P_{R}f+\text{c.c.} \,\,.
\end{equation}
The Yukawas $\lambda_{L,R}$ in general contain a $CP$-violating phase,
\begin{align}
  &\lambda_{L}\equiv\left|\lambda_{L}\right|e^{i\varphi/2}\,,& &\lambda_{R}\equiv\left|\lambda_{R}\right|e^{-i\varphi/2} \,\,.
\end{align}
There is a nonzero mixing angle $\alpha$ between the scalar mass and chiral eigenstates
\begin{equation}
  \left(\begin{array}{c}
    \widetilde{f}_{1} \\ \widetilde{f}_{2}
    \end{array}\right)
  =\left(
  \begin{array}{cc}
    \cos\alpha & -\sin\alpha \\
    \sin\alpha & \cos\alpha
  \end{array}\right)\left(
  \begin{array}{c}
    \widetilde{f}_{L} \\ \widetilde{f}_{R}
  \end{array}\right)\,.
\end{equation}
In the following, we denote the two scalar mass eigenvalues as $m_{\widetilde{f}_{1}}$ and $m_{\widetilde{f}_{2}}$. Our model thus has the following free parameters:
\begin{itemize}
\item the four masses $m_{\chi}$, $m_{\widetilde{f}_{1}}$, $m_{\widetilde{f}_{2}}$ and $m_{f}$. It is more convenient to use the following variables to represent the mass parameters:
\begin{align}
	&\mu_{1} =\frac{m^{2}_{\widetilde{f}_{1}}}{m_{\chi}^{2}}\,,& &\mu_{2}=\frac{m^{2}_{\widetilde{f}_{2}}}{m_{\chi}^{2}}\,,& &\delta=\frac{m_{f}^{2}}{m_{\chi}^{2}}\,.
\end{align}
%
If $f$ is a SM lepton, dipole moment constraints require that $f$ be either $\mu$ or $\tau$ \cite{Fukushima:2014yia,Fukushima:2013efa}.
\item the coupling constants $|\lambda_{L,R}|$, the $CP$-violation phase $\varphi$, and the scalar mixing angle $\alpha$. 
\end{itemize}

We briefly describe the supersymmetric limit of our simplified model. In the limit of the Minimal Supersymmetric Standard Model (MSSM), bino DM couples to one generation of light sleptons, and we have $|\lambda_{L}|=\sqrt{2}g|Y_{L}|$ and $|\lambda_{R}|=\sqrt{2}g|Y_{R}|$, where $g$ is the electroweak coupling constant, $|Y_{L}|=1/2$, and $|Y_{R}|=1$. {The mass squared matrix of the slepton in the chiral basis is
\begin{align}
-{\cal L} =
\begin{pmatrix}
\widetilde{f}_L^* & \widetilde{f}_R^*
\end{pmatrix}
\begin{pmatrix}
m_{\widetilde{f}_L}^2 & m_{\widetilde{f}_{LR}}^2\\
m_{\widetilde{f}_{LR}}^{2} & m_{\widetilde{f}_R}^2
\end{pmatrix}
\begin{pmatrix}
\widetilde{f}_L\\
\widetilde{f}_R
\end{pmatrix}\,.\label{massmatrix}
\end{align}
The relevant expressions for the matrix entries are:
\begin{subequations}\label{m}
\begin{align}
m_{\widetilde{f}_L}^2 &= m_{\widetilde{L}}^2 + m_Z^2\cos2\beta\bigl(-1/2+\sin^2\theta_w\bigr) + m_f^2\,, \label{mLL}\\
m_{\widetilde{f}_R}^2 &= m_{\widetilde{E}}^2 - m_Z^2\cos2\beta\sin^2\theta_w + m_f^2\,, \label{mRR}\\
m_{\widetilde{f}_{LR}}^2 &= m_f\bigl(A_f - \mu \tan\beta\bigr)\,.\label{mLR}
\end{align}
\end{subequations}
The mixing angle $\alpha$ is obtained as:
\begin{subequations}\label{theta2}
\begin{align}
\sin \alpha & = \frac{m_{\widetilde{f}_{LR}}^2}{\sqrt{(m_{\widetilde{f}_2}^2-m_{\widetilde{f}_L}^2)^2+(m_{\widetilde{f}_{LR}}^2)^2}},\\
\cos \alpha & = \frac{m_{\widetilde{f}_2}^2-m_{\widetilde{f}_L}^2}{\sqrt{(m_{\widetilde{f}_2}^2-m_{\widetilde{f}_L}^2)^2+(m_{\widetilde{f}_{LR}}^2)^2}}\,,
\end{align}
which leads to
\begin{align}
\tan \alpha & = \frac{m_f\bigl(A_f - \mu \tan\beta\bigr)}{m_{\widetilde{f}_2}^2-m_{\widetilde{f}_L}^2},
\label{theta3}
\end{align}
\end{subequations}
We note that for muons, using $|A_\mu - \mu \tan\beta | \sim 10^5$ GeV in Eq.~\eqref{theta3} yields $\tan \alpha \sim \mathcal{O}(1)$.}

The lepton anomalous dipole moments receive a new contribution from the vertex correction with the DM and scalars running in the loop. Thus current dipole moment measurements~\cite{Hanneke:2008tm,Baron:2013eja,Bennett:2006fi,Bennett:2008dy,Abdallah:2003xd,Inami:2002ah} are relevant. In the rest of the paper, {except for a discussion in Sec.~\ref{sec:g-2},} we will remain agnostic both about the lepton anomalous dipole moments as well as the relic density, and focus exclusively on the constraints coming from direct and indirect detection. This is in keeping with the spirit of simplified models, which tries to capture low energy constraints while keeping questions of high energy or early universe cosmology open. For example, the thermal relic density constraint depends heavily on the mechanism of thermal freezeout in the early Universe, which, given the ubiquity of moduli in UV complete frameworks, seems more and more unlikely \cite{Allahverdi:2013noa}. Non-thermal histories that can accommodate both overproducing and under-producing candidates have been studied in detail \cite{Dutta:2009uf, Kane:2015jia}. Similarly, dipole moment constraints in this class of models have been studied in detail, and we refer to \cite{Fukushima:2014yia,Kumar:2016cum} and references therein for details.

We note that the regions of parameter space that are most interesting for direct detection, with $\mu \leq 1.44$, are the ones where there are no constraints coming from colliders. Due to the high degeneracy of the mediator and DM, the leptons in the final state are soft and thus dilepton and trilepton searches cannot probe these regions ~\cite{ATLASslepton,CMSslepton}. Recent collider studies of compressed spectra show that these regions may be probed at high luminosity in weak boson fusion processes \cite{Dutta:2014jda, Dutta:2012xe, Delannoy:2013ata}. We thus take only LEP bounds~\cite{ALEPH,L3,DELPHI,OPAL} as our constraints and exclude charged scalars below $\sim 100\gev$.





\section{\label{sec:anapole}Direct Detection}

In this section, we discuss some general features of direct  detection for our simplified model, and the constraints we are going to use in Sec.~\ref{sec:result}.

\begin{figure*}[t]
    \subfloat[][$\mathcal{M}_{1}$]{\includegraphics[width=0.23\textwidth]{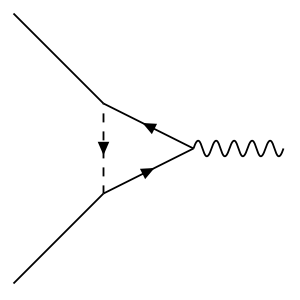}}
    \subfloat[][$\mathcal{M}_{2}$]{\includegraphics[width=0.23\textwidth]{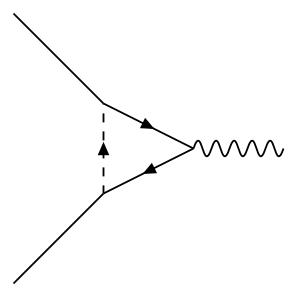}}
    \subfloat[][$\mathcal{M}_{3}$]{\includegraphics[width=0.23\textwidth]{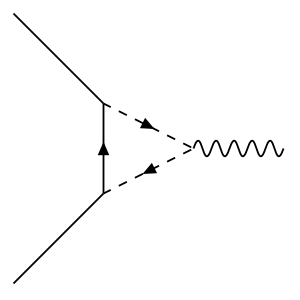}}
    \subfloat[][$\mathcal{M}_{4}$]{\includegraphics[width=0.23\textwidth]{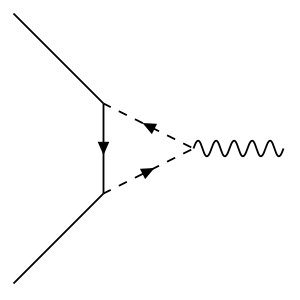}}
    \caption{Relevant Feynman diagrams for the anapole moment. $\mathcal{M}_{i}$ is the amplitude of each diagram, which are given in Appendix~\ref{sec:fullana}.}
    \label{fig:FeynAnapole}
\end{figure*}

In models with uncolored charged mediators, the DM interacts with nuclei only through the loop-induced electromagnetic moments. Moreover, the Majorana nature of our DM only allows a nonzero anapole moment. The relevant Feynman diagrams that contribute to the anapole moment are shown in Fig.~\ref{fig:FeynAnapole}. Since our DM is Majorana, we have also included the diagrams with the internal arrows reversed, which is equivalent to exchanging the two external fermions. 

If the incoming DM particle has momentum $p$ and the outgoing one has  momentum $p'$, the total off-shell amplitude given by Fig.~\ref{fig:FeynAnapole} is
\begin{equation}
\label{eq:Mmu}
  \mathcal{M}^{\mu}=i\mathcal{A}(q^{2})\overline{u}(p')\left(q^{2}\gamma^{\mu}-\slashed{q}q^{\mu}\right)\gamma^{5}u(p) \,\, ,
\end{equation}
where $q=p'-p$ is the momentum transfer and $\mathcal{A}(q^{2})$ is the anapole moment of the DM. Moreover, $\mathcal{A}(q^{2})$ can be expressed as
\begin{align}
\label{eq:A_alpha}
  \mathcal{A}(q^{2})&=e\left(\left|\lambda_{L}\right|^{2}\cos^{2}\alpha-\left|\lambda_{R}\right|^{2}\sin^{2}\alpha\right)X_{1}(q^{2})\nonumber\\
  &\quad +e\left(\left|\lambda_{L}\right|^{2}\sin^{2}\alpha-\left|\lambda_{R}\right|^{2}\cos^{2}\alpha\right)X_{2}(q^{2}) \,\,,
\end{align}
where $X_{1,2}$ is the result of three-point loop integrals. The derivation of the above two equations, together with the full form of $X_{i}$, will be given in Appendix~\ref{sec:fullana}. Noticeably, there is no $\varphi$ dependence in $\mathcal{A}$, because the amplitude $\mathcal{M}^{\mu}$ conserves $CP$. If both $p$ and $p'$ are on the same mass-shell, then the momentum transfer $q$ must be space-like, namely, $q^{2}<0$. 

In the limit $|q^{2}|\ll m_{f}^{2}$ and $|q^{2}|\ll m_{\widetilde{f}_{i}}^{2}$, $X_{i}$ has a simplified expression,
\begin{align}
X_{i}\xrightarrow{q^{2}=0}\frac{1}{96\pi^{2}m_{\chi}^{2}}\left[\frac{3\mu_{i}-3\delta+1}{\sqrt{\Delta_{i}}}\arctanh\left(\frac{\sqrt{\Delta_{i}}}{\mu_{i}+\delta-1}\right)-\frac{3}{2}\log\left(\frac{\mu_{i}}{\delta}\right)\right],
\end{align}
where $\Delta_{i}=(\mu_{i}-\delta-1)^{2}-4\delta$ and $\delta = m_f^2/m_\chi^2$. This limit applies approximately to DM direct detection for $f=\mu, \tau$. If the mediator is very heavy, $\mu_{i}\gg 1$, then $X_{i}$ indeed vanishes as $\mu_{i}^{-1}\log\mu_{i}$. On the other hand, if the mass gap between the scalar and the DM is small, the value of $X_{i}$ will be boosted; in the limit $(\mu_{i}-1)\sim\delta\ll 1$,
\begin{equation}
\label{eq:Xi}
  X_{i}\sim\frac{1}{96\pi^{2}m_{\chi}^{2}}\left[\frac{\pi}{\sqrt{\delta}}-\frac{3}{2}\log\frac{1}{\delta}\right].
\end{equation}
Therefore, for $f=\mu$ and $\tau$, our simplified model will give rise to a sizable anapole moment, which will lead to signals in direct detection experiments.

In the parameter space of our interest, one scalar mediator is quite degenerate with the DM while the other is heavy. Thus, we have $\mu_{1}\sim 1$ and $\mu_{2}\gg 1$ such that $X_{1}\gg X_{2}$. Then the ``blind spot'' in the parameter space is located around
\begin{equation*}
\tan\alpha\sim\left|\frac{\lambda_{L}}{\lambda_{R}}\right|\,,
\end{equation*}
where $\mathcal{A}$ is suppressed due to the lack of the contribution from $X_1$.

The DM mass range of interest here is $\mathcal{O}(100\gev)$, where the most sensitive constraints are drawn by Xenon-based experiments, such as LUX, XENON100~\cite{Aprile:2012nq}, PandaX~\cite{Tan:2016diz}, and future LZ, XENON1T~\cite{Aprile:2015uzo}, etc. The typical nuclear recoil event has energy of $\sim10$-$30\kev$, which corresponds to a DM-nucleus momentum transfer of
\begin{equation*}
\sqrt{|q^{2}|}\approx|\pmb{q}|\approx 50\sim 80\mev,
\end{equation*}
where $\pmb{q}$ is the three-momentum of the four-momentum $q$. In our model, the only DM-nucleus interaction is mediated by the anapole, which can be described by the following effective Lagrangian at small momentum transfer:
\begin{equation}
\label{eq:analag}
\mathcal{L}_{\text{DM-nucleus}}=\frac{i\mathcal{A}}{2}\overline{\chi}\gamma^{\mu}\gamma^{5}\chi\partial^{\nu}F_{\mu\nu}+eA_{\mu}J^{\mu},
\end{equation}
where $\mathcal{A}\equiv\mathcal{A}(0)$ and $J^{\mu}$ is the nuclear current operator. {The anapole moment $\mathcal{A}$ is real, since the combination $i\overline{\chi}\gamma^{\mu}\gamma^{5}\chi\partial^{\nu}F_{\mu\nu}$ is real.} This interaction conserves $CP$, but not $C$ and $P$ individually. Using the full expression given in Appendix~\ref{sec:fullana}, we have checked that by neglecting the $q^{2}$ dependence, we only introduce, at most, a $0.6\%$ error in the anapole moment for the models studied here. The differential cross section for an anapole DM (with speed $v^{2}\sim 10^{-6}$) scattering off a target nuclear electric field is~\cite{Kopp:2014tsa,DelNobile:2014eta,Ho:2012br,Gresham:2013mua}
\begin{equation}
\label{eq:dsigmadE}
\frac{d\sigma}{dE_{R}}=\alpha_{\text{em}}\mathcal{A}^{2}Z^{2}F^{2}_{E}(\pmb{q}^{2})\left[2m_{T}-\left(1+\frac{m_{T}}{m_{\chi}}\right)^{2}\frac{E_{R}}{v^{2}}\right] \,\,.
\end{equation}
In this equation, $m_{T}$ is the mass of the target nucleus (xenon for LUX), $E_{R}=\pmb{q}^{2}/(2m_T)$ is the nuclear recoil energy, and $Z$ is the nuclear charge. The nuclear electric form factor $F^{2}_{E}$ is taken as the Helm form factor~\cite{Helm:1956zz}. In principle, anapole DM also interacts with the nuclear magnetic field, but this contribution is negligible for xenon nuclei due to the smallness of the nuclear magnetic dipole moment~\cite{DelNobile:2014eta}. Eq.~\eqref{eq:dsigmadE} exhibits a different DM velocity dependence from the usual spin-independent (SI) one. As a result, the LUX SI constraint presented in~\cite{Akerib:2013tjd,Akerib:2015rjg} cannot be directly applied here. Instead, one needs to calculate the scattering rate using the standard halo model (SHM) and fit the LUX data at the event level following~\cite{Kopp:2014tsa}. For future LZ projections, the strategy we employ is as follows: We scale the constraint on the scattering cross section for a given DM mass from the LUX limit to the projected LZ limit, resulting in a scaling of the constraint on the anapole moment, from Eq.~\eqref{eq:dsigmadE}. This constraint on the anapole moment is then further translated into constraints on the parameters of the model Lagrangian. As a full event-level analysis has not yet been performed for the LUX 2016 results, we follow the same strategy when estimating the LUX 2016 bounds on the parameter space. Clearly, a more careful analysis of the LUX 2016 datasets, which is beyond the scope of the present work, will sharpen these constraints.

We turn next to indirect detection.


\section{\label{sec:IB}Indirect Detection}

In this Section, we outline the indirect detection constraints on our model. We will only focus on gamma-ray searches. We note that AMS bounds~\cite{Bergstrom:2013jra} may also come into play, but we do not consider them here due to uncertainties related to the astrophysical background and propagation of charged particles.

We first begin with a discussion of the chiral suppression of the annihilation cross section in this class of models, and how it is lifted through either internal bremsstrahlung (IB) processes or non-zero mixing of the two mediators. We then go on to discuss the $Fermi$-LAT constraints on our model.

In the $\alpha\rightarrow 0$ and $m_{f}\rightarrow 0$ limit, the chiral symmetry $f\rightarrow\exp(i\theta\gamma^{5})f$ forbids the $s$-wave two-body annihilation $\chi\chi\rightarrow f\bar{f}$ in the current era. The reason is as follows. DM particles at the current era typically have relative velocity $v\rightarrow 0$ such that the Majorana nature requires the initial state be in the total angular momentum $J=0$ state. Then the conservation of angular momentum requires the final state fermion anti-fermion pair be of the same helicity, which, for $m_{f}\rightarrow 0$, can be achieved only if the fermion is left-handed and the anti-fermion is right-handed. Then this amplitude is not invariant under a chiral symmetry transformation and is thus forbidden. Since a small mass $m_{f}$ minimally violates the chiral symmetry, the annihilation cross section must scale as $(m_{f}/m_{\chi})^{2}$.  That is, it is chirally suppressed. 

A finite $\alpha$ deviates from the minimal violation explicitly and thus enables an unsuppressed $s$-wave cross section, making the DM annihilation signal large enough to be potentially observed\footnote{The $\varphi$ dependence of a generic amplitude is always chirally-suppressed, being proportional to $(m_{f}/m_{\chi})\sin 2\alpha$. The reason is that at $m_{f}=0$ ($\alpha=0$ or $\pi/2$), the $\varphi$ dependence can be absorbed by a chiral rotation of the fermion (scalar).}. A left-right scalar mixing in our simplified model \eqref{eq:Lint} thus enables an unsuppressed $s$-wave annihilation~\cite{Fukushima:2014yia,Kumar:2016cum}. Chiral suppression can also be lifted by introducing one more photon in the final state, which modifies the condition of angular momentum conservation~\cite{Bringmann:2007nk,Bergstrom:1989jr}.  When considering IB in our model, both mechanisms are encoded.

We present briefly the IB calculation, which gives the indirect detection signal, and refer to~\cite{Kumar:2016cum} for a dedicated study. In the $s$-wave limit, the total IB amplitude can be written as the sum of three gauge invariant sub-amplitudes,
\begin{equation}
\label{eq:AIB}
  \mathcal{A}_{\text{IB}}=\frac{ie}{2}\left[\frac{\overline{u}(k_{1})\gamma^{5}v(k_{2})}{2m_{\chi}}\right]\left(\mathcal{A}_{\text{vb}}+\mathcal{A}_{\text{mix}}+\mathcal{A}_{m_{f}}\right),
\end{equation}
all of which have clear physical meanings. Here $k_{1}=k_{2}=k$ are the momenta of the initial DM particles, while $u$ and $v$ are standard spinor wavefunctions following the definition of~\cite{Peskin}. The full analytic expressions for these three sub-amplitudes are given in Appendix \ref{sec:fullIB}. First, $\mathcal{A}_{\text{vb}}$ is an intrinsic unsuppressed $s$-wave amplitude. This amplitude is contributed by the final states in which the fermion and anti-fermion have opposite helicities, made possible by the vector boson emission, which lifts the chiral suppression. Thus it survives even in the limit $\alpha=0$, when the minimal chiral symmetry violation is restored. The contribution of $\mathcal{A}_{\text{vb}}$ features a line like photon spectrum (if at least one scalar is very degenerate with the DM) with the peak around the kinematic cut-off $E_{\gamma}\approx m_{\chi}$~\cite{Bringmann:2007nk}:
\begin{align}
\frac{d(\sigma v)_{\text{vb}}}{dx}=\sum_{i=1,2}\frac{\alpha_{\text{em}}\lambda_{i}^{4}(1-x)}{64\pi^{2}m_{\chi}^{2}}&\left[\frac{4x}{(1+\mu_{i})(1+\mu_{i}-2x)}-\frac{2x}{(1+\mu_{i}-2x)^{2}}\right.\nonumber\\
&\;\left.-\frac{(1+\mu_{i})(1+\mu_{i}-2x)}{(1+\mu_{i}-x)^{3}}\log\frac{1+\mu_{i}}{1+\mu_{i}-2x}\right],
\end{align}
where $x=E_{\gamma}/m_{\chi}$ is the photon energy fraction, and
\begin{align}
&\lambda_{1}=|\lambda_{L}|^{2}\cos^{2}\alpha-|\lambda_{R}|^{2}\sin^{2}\alpha\,,& &\lambda_{2}=|\lambda_{L}|^{2}\sin^{2}\alpha-|\lambda_{R}|^{2}\cos^{2}\alpha.
\end{align}
Nonzero scalar mixing angle $\alpha$ opens another unsuppressed amplitude, $\mathcal{A}_{\text{mix}}\propto\sin(2\alpha)$. Unlike $\mathcal{A}_{\text{vb}}$, this term is induced by an explicit deviation from the minimal violation of the chiral symmetry by introducing the scalar mixing. At finite $\alpha$, $\mathcal{A}_{\text{mix}}$ is divergent in both the soft and collinear limit when $m_{f}\rightarrow 0$, which gives the dominant contribution to $\mathcal{A}_{\text{IB}}$:
\begin{align}
&\frac{d(\sigma v)_{\text{IB}}}{dx}\approx\frac{\alpha_{\text{em}}}{\pi}\frac{x^{2}-2x+2}{x}\log\left[\frac{s(1-x)}{m_{f}^{2}}\right]\times(\sigma v)_{f\bar{f}}& &\text{(for finite $\alpha$)},
\end{align}
where $s=4m_{\chi}^{2}$, and $(\sigma v)_{f\bar{f}}$ is the unsuppressed $s$-wave two-body annihilation cross section. {The total cross section thus has to be modulated by a Sudakov double log factor.} It also washes out the peak like feature of $\mathcal{A}_{\text{vb}}$ when $\alpha\gtrsim\pi/100$ for our typical benchmark models~\cite{Kumar:2016cum}. Finally, $\mathcal{A}_{m_{f}}\propto m_{f}$ is the chirally suppressed part. It has the same $\alpha$ dependence as $\mathcal{A}_{\text{vb}}$ but the similar spinor chain structure as $\mathcal{A}_{\text{mix}}$. Consequently, it also survives at $\alpha=0$, which reflects the effect of the minimal chiral symmetry violation. On the other hand, like $\mathcal{A}_{\text{mix}}$, it is divergent in the soft limit, but such a divergence is very mild due to the chiral suppression.

To sum up, our simplified model \eqref{eq:Lint} incorporates two mechanisms to lift the chiral suppression on the IB cross section. At $\alpha\rightarrow 0$, the photon emission enables the same-helicity fermion anti-fermion pair in the final state, which leads to a peak-like spectrum. The peak is prominent if at least one scalar mass is very degenerate with the DM mass. At finite $\alpha$, the deviation from the minimal chiral symmetry violation lifts the chiral suppression, but leads to a flat spectrum instead. The lesson for  indirect detection is that we need to use line signal searches to constrain the no mixing case with degenerate spectrum, namely, $\mu_1=1.01$ and $\mu_1=1.1$, but continuum searches to constrain the finite mixing case and nondegenerate spectrum ($\mu_1=1.44$). {For $\mu_{1}\lesssim 1.1$, the position of the sharp peak in the IB spectrum can be found from the equation
  \begin{equation*}
    \frac{d}{dx}\left[\frac{d(\sigma v)_{\text{IB}}}{dx}\right]\approx\frac{d}{dx}\left[\frac{d(\sigma v)_{\text{vb}}}{dx}\right]=0\,.
  \end{equation*}
We find that the peak always sits at $E_{\gamma}\gtrsim 0.91 m_{\chi}$ for $\mu_{1}\lesssim 1.1$. When applying the line constraint, we neglect this small difference since the $Fermi$-LAT constraint does not change much in this range.}


Now we discuss the constraints coming from indirect detection. As discussed above, both limits drawn by line and continuum searches might be relevant. In the DM mass range of interest, the most sensitive results come from dwarf galaxies. In the future, these limits can be improved by GAMMA400~\cite{G400} and HERD~\cite{HERD}, which are designed to have better sensitivity and energy resolution. 

For the line search limit that constrains the model at $\alpha=0$ or $\pi/2$ with $\mu_{1}\lesssim 1.1$, we use the PASS 8 analysis result of $5.8$ years data~\cite{Ackermann:2015lka} of $Fermi$-LAT. At $m_{\chi}=100\gev$, the bound on the annihilation cross section is about $4.5\times10^{-28}\text{cm}^{3}/\text{s}$. The uncertainty of this limit spans about one order of magnitude. {The one-loop suppressed processes $\chi\chi\rightarrow\gamma\gamma$ and $\gamma Z$ also produce line signals, but for $m_{\chi}\gtrsim 100\gev$, the mass range of interest here, the cross sections for these processes are only about $10^{-30}\sim 10^{-31}\text{cm}^{3}/\text{s}$ with SUSY couplings. Thus the IB signal always dominates, which gives a total cross section of about $10^{-28}\sim 10^{-29}\text{cm}^{3}/\text{s}$, also with SUSY couplings.}

For the continuum spectrum search limit, we use the one on the particle physics factor $\Phi_{\text{PP}}$ given in~\cite{GeringerSameth:2011iw}:
\begin{equation*}
\Phi_{\text{PP}}=5.0^{+4.3}_{-4.5}\times 10^{-30}\text{cm}^{3}\text{s}^{-1}\text{GeV}^{-2},
\end{equation*}
from which the limit on the IB cross section can be inferred as
\begin{equation}
(\sigma v)_{\text{IB}}=(8\pi m_{\chi}^{2})\Phi_{\text{PP}}=\left(\frac{m_{\chi}}{100\gev}\right)^{2}\times 1.26^{+1.08}_{-1.13}\times 10^{-24}\text{cm}^{3}/\text{s} \,\,.
\end{equation}
Although this is not a very recent analysis, it is more directly applicable to our case than the latest $Fermi$-LAT analysis (for example, \cite{Ackermann:2015zua}) since the limit is drawn without assuming a spectrum. In common practice, one needs to first generate a photon spectrum from the decay chain of the final state (for example, simulated by \texttt{PYTHIA}~\cite{Sjostrand:2014zea}) and then fit it with the observed spectrum. In our work, the spectrum is approximated by an analytic calculation of the IB, which can in principle be different from the \texttt{PYTHIA} simulation. {We note that for the $\mu^{+}\mu^{-}$ final state IB is the dominant contribution and the difference is relatively small. In this case, the bounds we obtain from $\Phi_{\text{PP}}$ are somewhat conservative, but are robust and spectrum independent. If we neglect the spectrum difference and directly apply the $Fermi$-LAT result, the improvement is about a factor of $6$ at $m_{\chi}=100\gev$ for $(\sigma v)_{\mu^{+}\mu^{-}}$, which is about a factor of $\sqrt[4]{6}\approx 1.56$ improvement in the constraint on the coupling $\lambda_{L,R}$. We would like to encourage an updated spectrum-independent analysis of the $Fermi$-LAT signal from dwarf galaxies.}

\section{\label{sec:result}Results}



In this section, we describe the constraints on the parameter space of our model coming from direct and indirect detection.  We present our results for these constraints as contours in the following planes: $(\alpha,\lambda)$, $(m_\chi,\lambda)$, and $(\lambda_L,\lambda_R)$. {All the results  presented are for the $\mu$ channel. For the $\tau$ final state, the direct detection constraints are weaker, since $\mathcal{A}\sim m_{f}^{-1}$ according to Eq.~\eqref{eq:Xi}. However, the indirect bounds from $Fermi$-LAT are stronger for annihilation to taus~\cite{Ackermann:2015zua}. We also emphasize that the $|q^{2}|\ll m_{f}^{2}$ approximations presented in Sec.~\ref{sec:anapole} are used only  to give a qualitative understanding of the behavior of the anapole moment, while the results in this section are calculated from the full analytic expression given in Appendix~\ref{sec:fullana}.}

\subsection{\texorpdfstring{$(\alpha, \lambda)$}{(alpha,lambda)} Planes}

\begin{figure}[ht]
  \centering
\includegraphics[width=\textwidth]{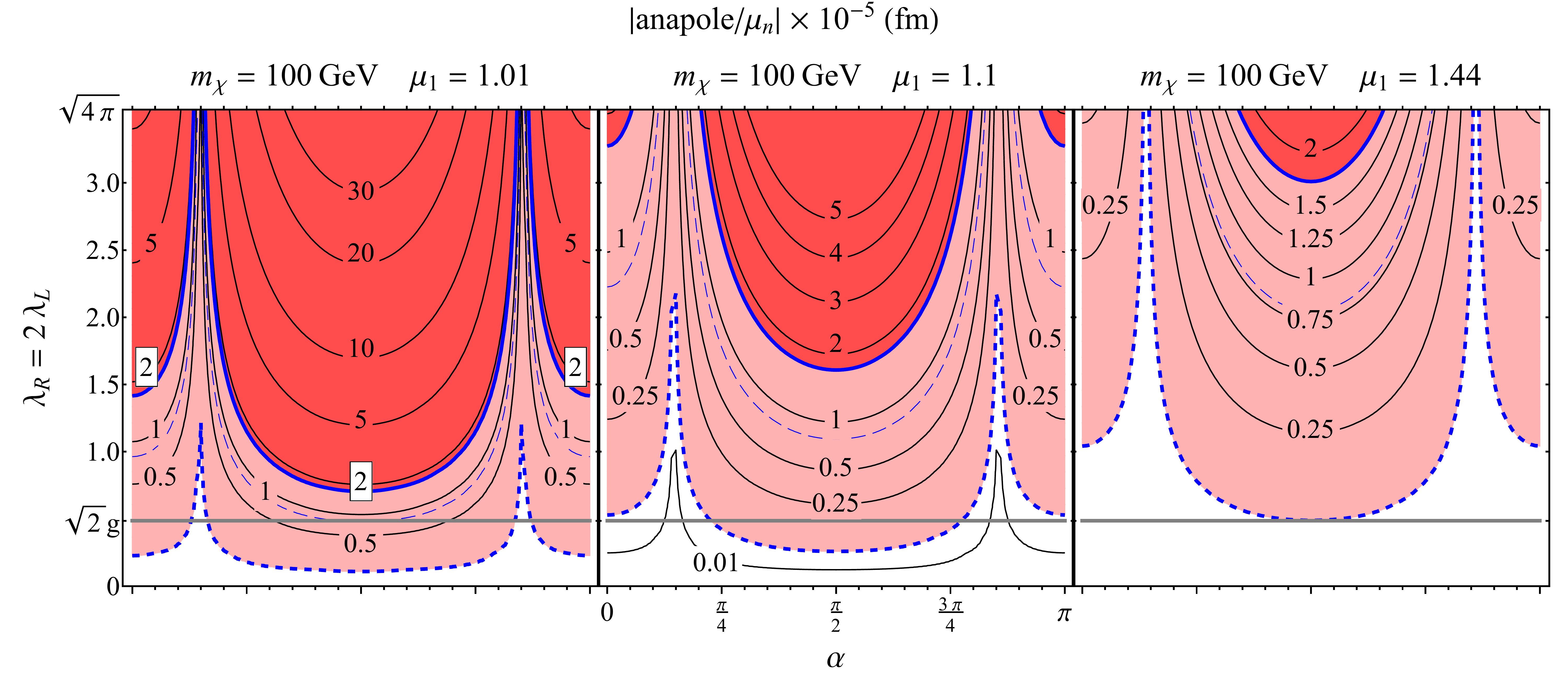}\\
\includegraphics[width=\textwidth]{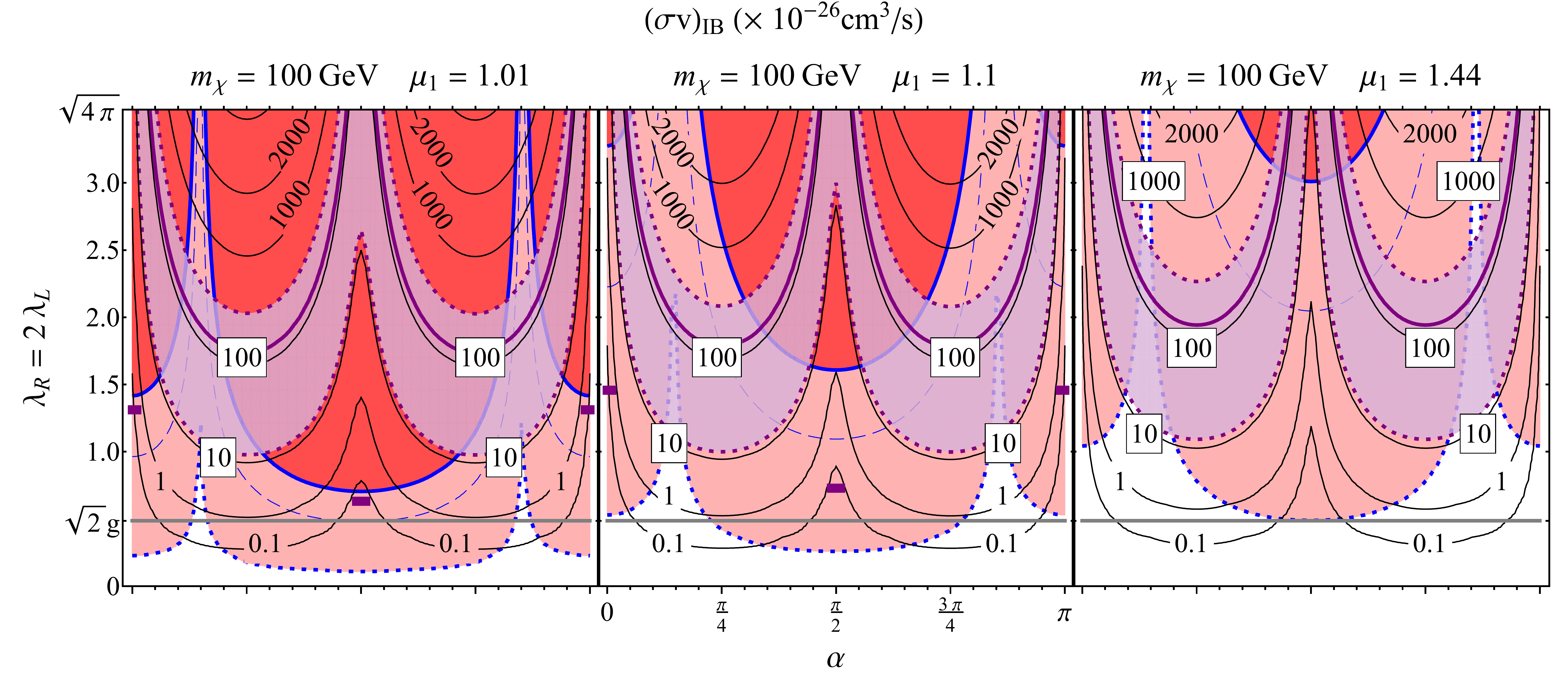}
\caption{$({\alpha},{\lambda})$ with $m_{\chi} = 100$ GeV: We show plots of $\lambda_R = 2 \lambda_L$ versus $\alpha$ for $\mu = 1.01$ (left), $\mu = 1.10$ (center), and $\mu = 1.44$ (right). In the first row, the contours correspond to values of $|\mathcal{A}/\mu_N| \times 10^{-5}$ fm. The solid (dashed) blue lines correspond to LUX 2014 (future LZ) limits on the DM SI scattering cross section.  In addition, the most recent LUX 2016 constraint is estimated as a thin dashed blue contour.  The gray horizontal line corresponds to the SUSY value of couplings. In the second row, the red shaded regions and blue contours for LUX sensitivity remain the same, while additional black contours correspond to values of the IB cross section $(\sigma v)_{\text{IB}} \times 10^{-26}$ cm$^3$s$^{-1}$. The solid purple contour corresponds to the central value of the limits placed by $Fermi$-LAT on the DM annihilation cross section coming from dwarf galaxies, while the dashed purple contours correspond to $95\%$ CL interval. The thick purple horizontal line segments at $\alpha = 0, \pi/2,$ and $\pi$ correspond to limits from $Fermi$-LAT line searches. }
  \label{Yalpha100}
\end{figure}

\begin{figure}[th]
\centering
\includegraphics[width=\textwidth]{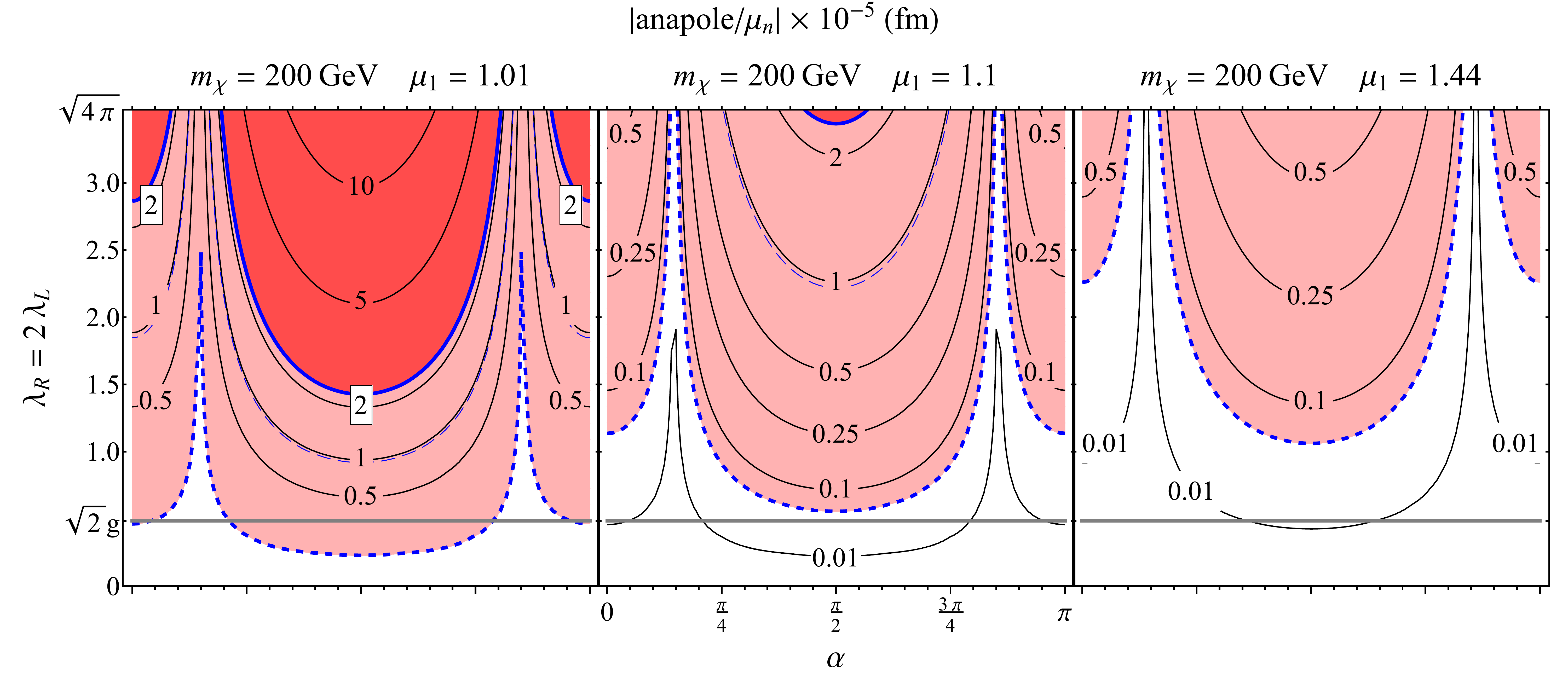}\\
\includegraphics[width=\textwidth]{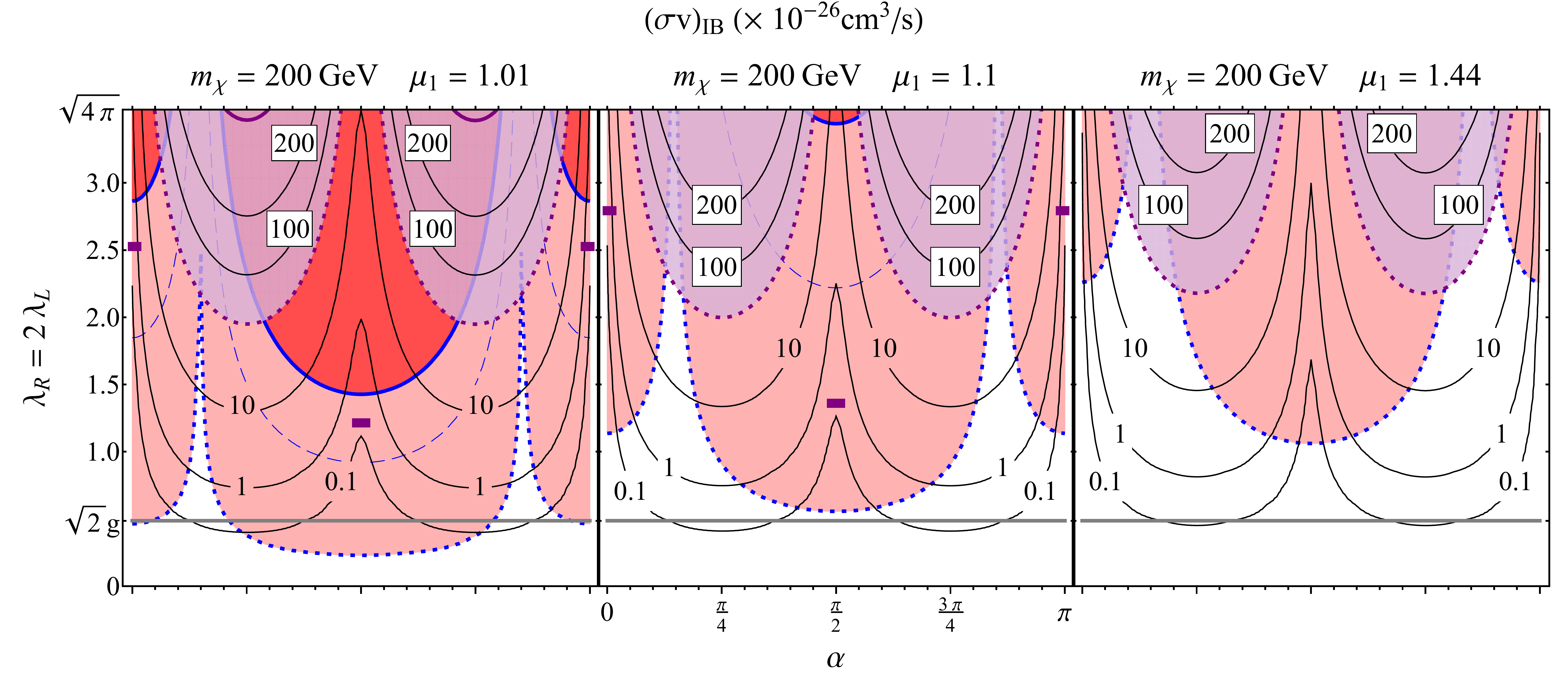}
\caption{\label{Yalpha200} $({\alpha},{\lambda})$ with $m_{\chi} = 200$ GeV: We show plots of $\lambda_R = 2 \lambda_L$ versus $\alpha$ for $\mu = 1.01$ (left), $\mu = 1.10$ (center), and $\mu = 1.44$ (right). In the first row, the contours correspond to values of $|\mathcal{A}/\mu_N| \times 10^{-5}$ fm. The solid (dashed) blue lines correspond to LUX 2014 (future LZ) limits on the DM SI scattering cross section.  In addition, the most recent LUX 2016 constraint is estimated as a thin dashed blue contour.  The gray horizontal line corresponds to the SUSY value of couplings. In the second row, the red shaded regions and blue contours for LUX sensitivity remain the same, while additional black contours correspond to values of the IB cross section $(\sigma v)_{\text{IB}} \times 10^{-26}$ cm$^3$s$^{-1}$. The solid purple contour corresponds to the central value of the limits placed by $Fermi$-LAT on the DM annihilation cross section coming from dwarf galaxies, while the dashed purple contours correspond to $95\%$ CL interval. The thick purple horizontal line segments at $\alpha = 0, \pi/2,$ and $\pi$ correspond to limits from $Fermi$-LAT line searches. }
\end{figure}

Our results for the LUX/LZ and $Fermi$-LAT sensitivities to our models are presented in the $(\alpha,\lambda)$ planes in Figs.~\ref{Yalpha100} and \ref{Yalpha200}.

In Figure~\ref{Yalpha100}, we present our results in the $(\lambda,\alpha)$ plane. We show plots of $\lambda_R = 2 \lambda_L$ versus $\alpha$, for three different values of $\mu$: $\mu = 1.01$ (left), $\mu = 1.10$ (center), and $\mu = 1.44$ (right). In the first row, the contours correspond to values of $|\mathcal{A}/\mu_N| \times 10^{-5}$ fm. The solid (dashed) blue lines correspond to LUX 2014 (future LZ) limits on the DM SI scattering cross section.  In addition, the most recent LUX 2016 constraint is estimated as a thin dashed blue contour using the procedure described for the projected LZ sensitivity.  As a full event-level analysis has not yet been performed for the LUX 2016 results, we focus the following discussion on the 2014 results. The gray horizontal line corresponds to the SUSY value of couplings. In the second row, the red shaded regions and blue contours for LUX sensitivity remain the same, while additional black contours correspond to values of the IB cross section $(\sigma v)_{\text{IB}} \times 10^{-26}$ cm$^3$s$^{-1}$. The solid purple contour corresponds to the central value of the limits placed by $Fermi$-LAT on the DM annihilation cross section coming from dwarf galaxies, while the dashed purple contours correspond to $95\%$ CL interval. The thick purple horizontal line segments at $\alpha = 0, \pi/2,$ and $\pi$ correspond to limits from $Fermi$-LAT line searches. 

Indirect detection places complementary constraints on the parameter space. From the second row of Figure \ref{Yalpha100}, it is clear that regions near $\alpha = \pi/8, 7\pi/8$ {(the ``blind spot'' region)} are already being probed by $Fermi$-LAT continuum searches, which place stronger limits than LUX bounds in these regions of parameter space. Conversely, LUX bounds are stronger than bounds from $Fermi$-LAT in the region near $\alpha = \pi/2$. Limits from line searches, which are applicable at $\alpha = 0, \pi/2, \pi$ are comparable with LUX bounds. For larger values of $\mu$, the bounds from indirect detection remain approximately unchanged, while those from direct detection degrade significantly.

We now describe the ``blind spots" in the parameter space, where the anapole moment nearly vanishes, obvious in Fig.~\ref{Yalpha100} as the places where direct detection constraints are weakest, near $\alpha = \pi/8, 7\pi/8$, from Eq.~\eqref{eq:A_alpha}. Even in the best case scenario of $\mu = 1.01$, LUX bounds do not probe these regions, although they will be probed by LZ. It is clear from the left plot in the first row that LUX bounds also do not probe the SUSY limit, while LZ will probe most of the SUSY limit for $\mu = 1.01$. For larger values of $\mu$, the prospects for direct detection are considerably weaker. For  $\mu = 1.44$, even future LZ bounds will barely start to constrain the SUSY parameter space. 

In Figure~\ref{Yalpha200}, we repeat the results of Figure~\ref{Yalpha100}, but for $m_\chi = 200$ GeV. While the constraints show the same general features, there is an appreciable deterioration of the reach due to the dependence on DM mass in Eq.~\eqref{eq:A_alpha}. In fact, now the future LZ projections barely touch the supersymmetric limit for $\mu = 1.10$, and the SUSY limit is completely unconstrained for $\mu = 1.44$. The constraints from $Fermi$-LAT show a similar degeneration. In both cases, we see that the bounds from $Fermi$-LAT continuum searches are comparable with LUX. Future LZ projections, however, go further in probing the parameter space than the current indirect detection limits.



\subsection{\texorpdfstring{$(m_\chi,\lambda)$}{(m,lambda)} Planes}

\begin{figure}[t]
	\centering
	\includegraphics[width=\textwidth]{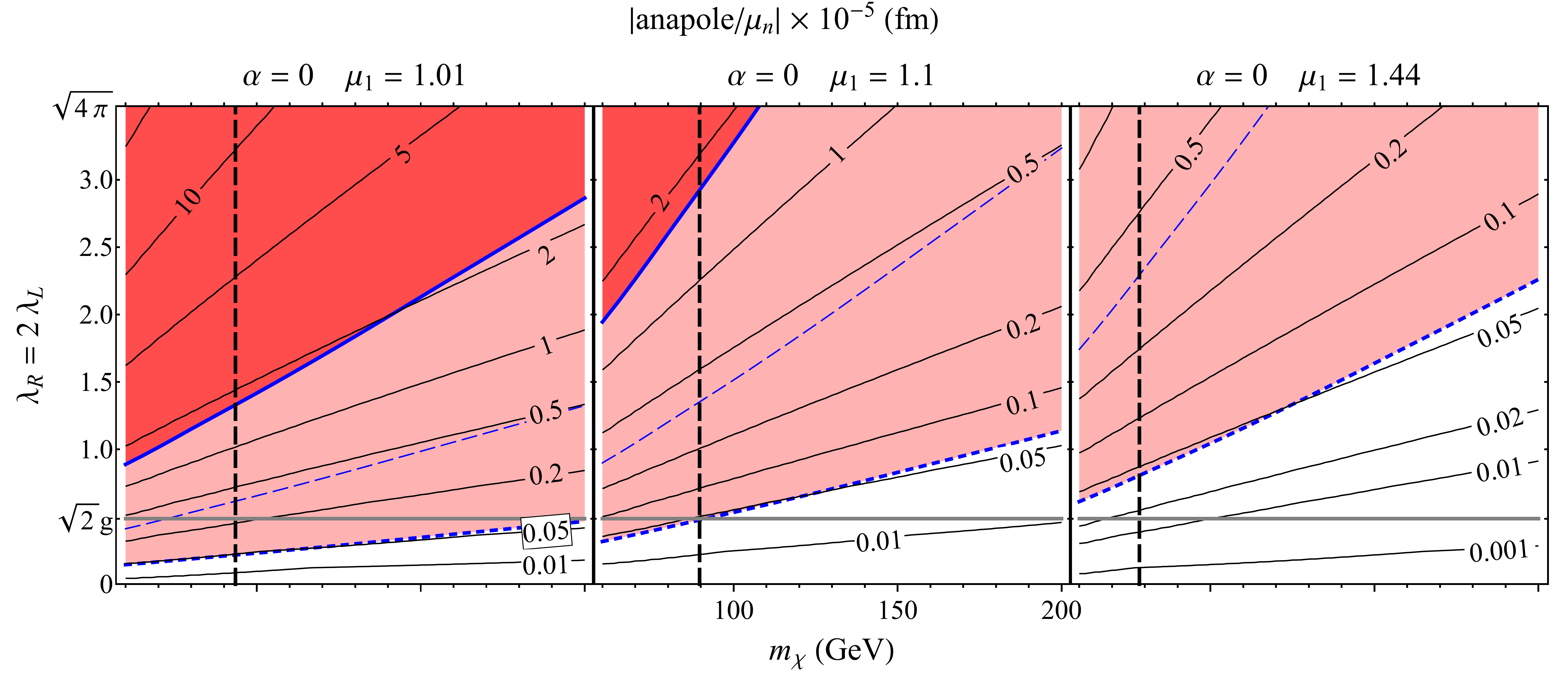}\\
	\includegraphics[width=\textwidth]{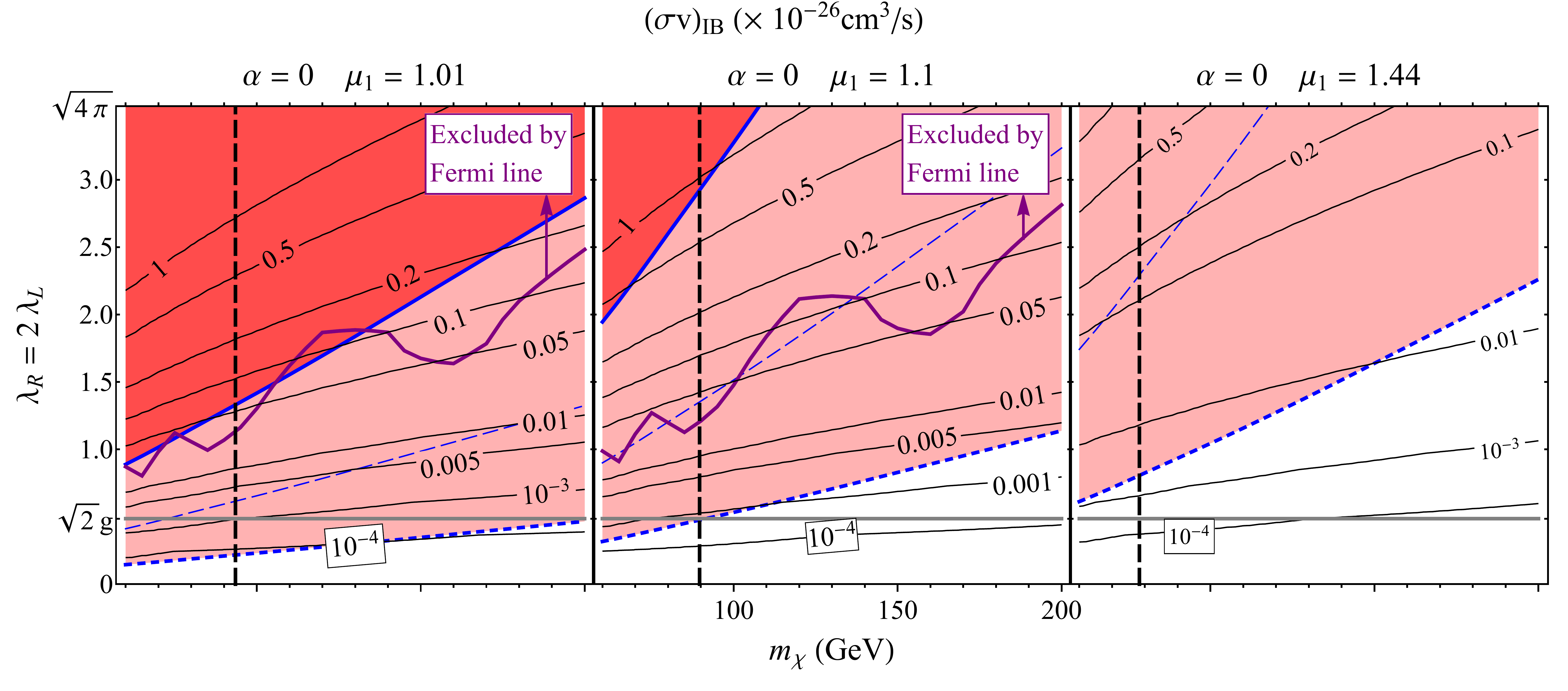}
\caption{$({m_\chi},{\lambda})$ for ${\alpha = 0}$: We show plots of $\lambda_R = 2 \lambda_L$ versus $m_\chi$ for $\mu = 1.01$ (left), $\mu = 1.10$ (center), and $\mu = 1.44$ (right). In the first row, the contours correspond to values of $|\mathcal{A}/\mu_N| \times 10^{-5}$ fm. {The vertical black dashed line is the LEP limit~\cite{ALEPH,L3,DELPHI,OPAL} on the mass of the charged scalar for the $\mu$ channel.} The solid (dashed) blue lines correspond to LUX 2014 (future LZ) limits on the DM SI scattering cross section.  In addition, the most recent LUX 2016 constraint is estimated as a thin dashed blue contour.  The gray horizontal line corresponds to the SUSY value of couplings. In the second row, the red shaded regions and blue contours for LUX sensitivity remain the same, while additional black contours correspond to values of the IB cross section $(\sigma v)_{\text{IB}} \times 10^{-26}$ cm$^3$s$^{-1}$. The solid purple contour corresponds to the central value of the limits placed by $Fermi$-LAT on the DM annihilation cross section coming from dwarf galaxies, while the dashed purple contours correspond to $95\%$ CL interval. The thick purple horizontal line segments at $\alpha = 0, \pi/2,$ and $\pi$ correspond to limits from $Fermi$-LAT line searches.}
  \label{Ymassalpha0andpiover4}
\end{figure}

\begin{figure}[t]
	\centering
	\includegraphics[width=\textwidth]{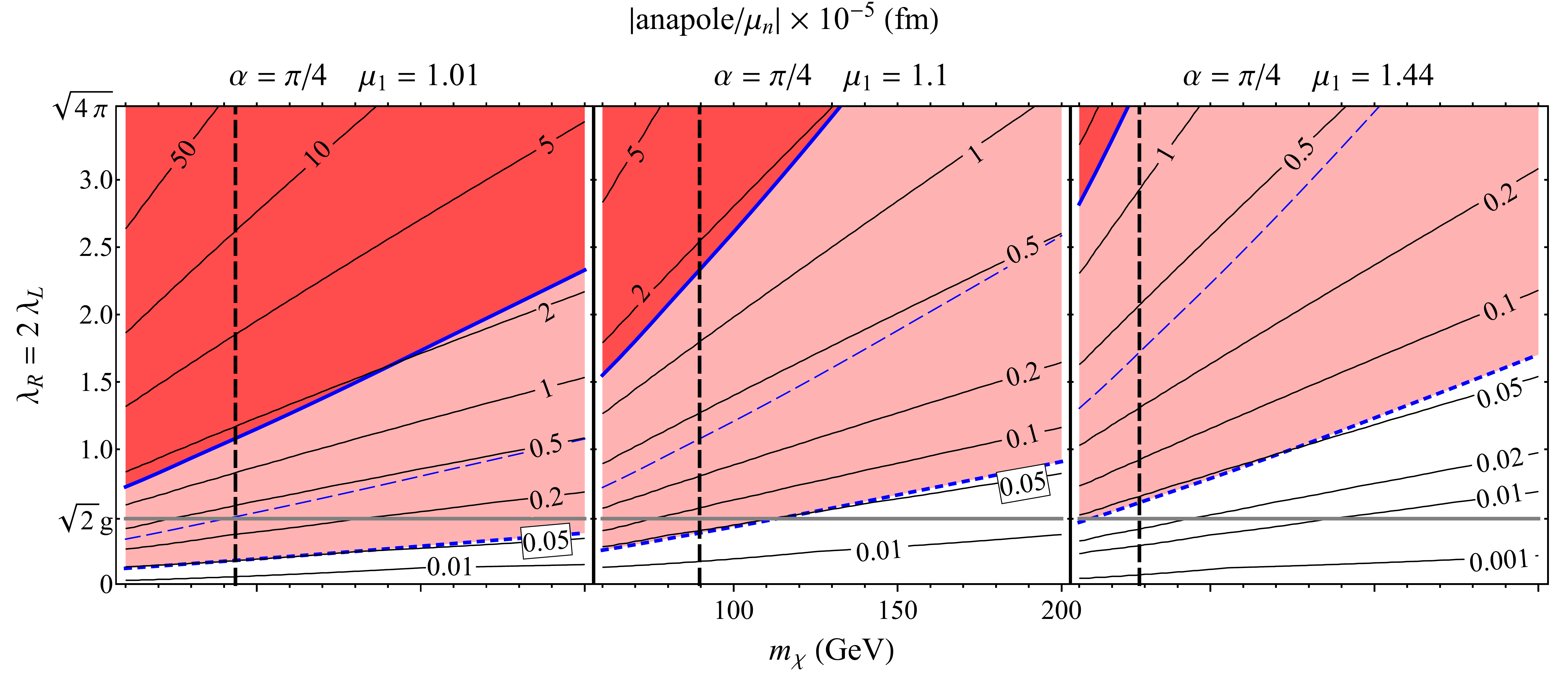}\\
	\includegraphics[width=\textwidth]{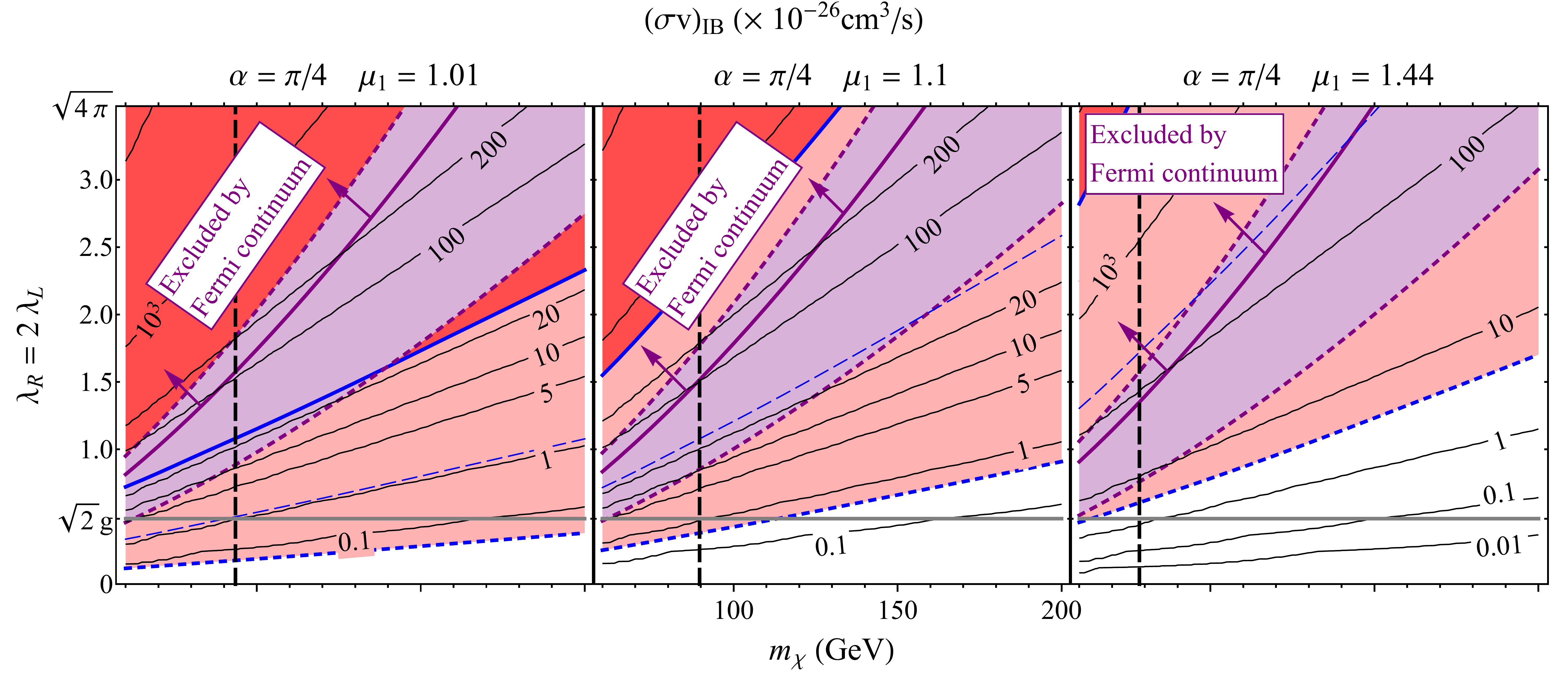}
	\caption{$({m_\chi},{\lambda})$ for ${\alpha = \pi/4}$: We show plots of $\lambda_R = 2 \lambda_L$ versus $m_\chi$ for $\mu = 1.01$ (left), $\mu = 1.10$ (center), and $\mu = 1.44$ (right). In the first row, the contours correspond to values of $|\mathcal{A}/\mu_N| \times 10^{-5}$ fm. The vertical black dashed line is the LEP limit on the mass of the charged scalar for the $\mu$ channel.  The solid (dashed) blue lines correspond to LUX 2014 (future LZ) limits on the DM SI scattering cross section.  In addition, the most recent LUX 2016 constraint is estimated as a thin dashed blue contour.  The gray horizontal line corresponds to the SUSY value of couplings. In the second row, the red shaded regions and blue contours for LUX sensitivity remain the same, while additional black contours correspond to values of the IB cross section $(\sigma v)_{\text{IB}} \times 10^{-26}$ cm$^3$s$^{-1}$. The solid purple contour corresponds to the central value of the limits placed by $Fermi$-LAT on the DM annihilation cross section coming from dwarf galaxies, while the dashed purple contours correspond to $95\%$ CL interval. The thick purple horizontal line segments at $\alpha = 0, \pi/2,$ and $\pi$ correspond to limits from $Fermi$-LAT line searches. In all cases, the spectrum shows a continuum feature.}
	  \label{Ymassalpha0andpiover4b}
\end{figure}

\begin{figure}[t]
  \centering
  	\includegraphics[width=\textwidth]{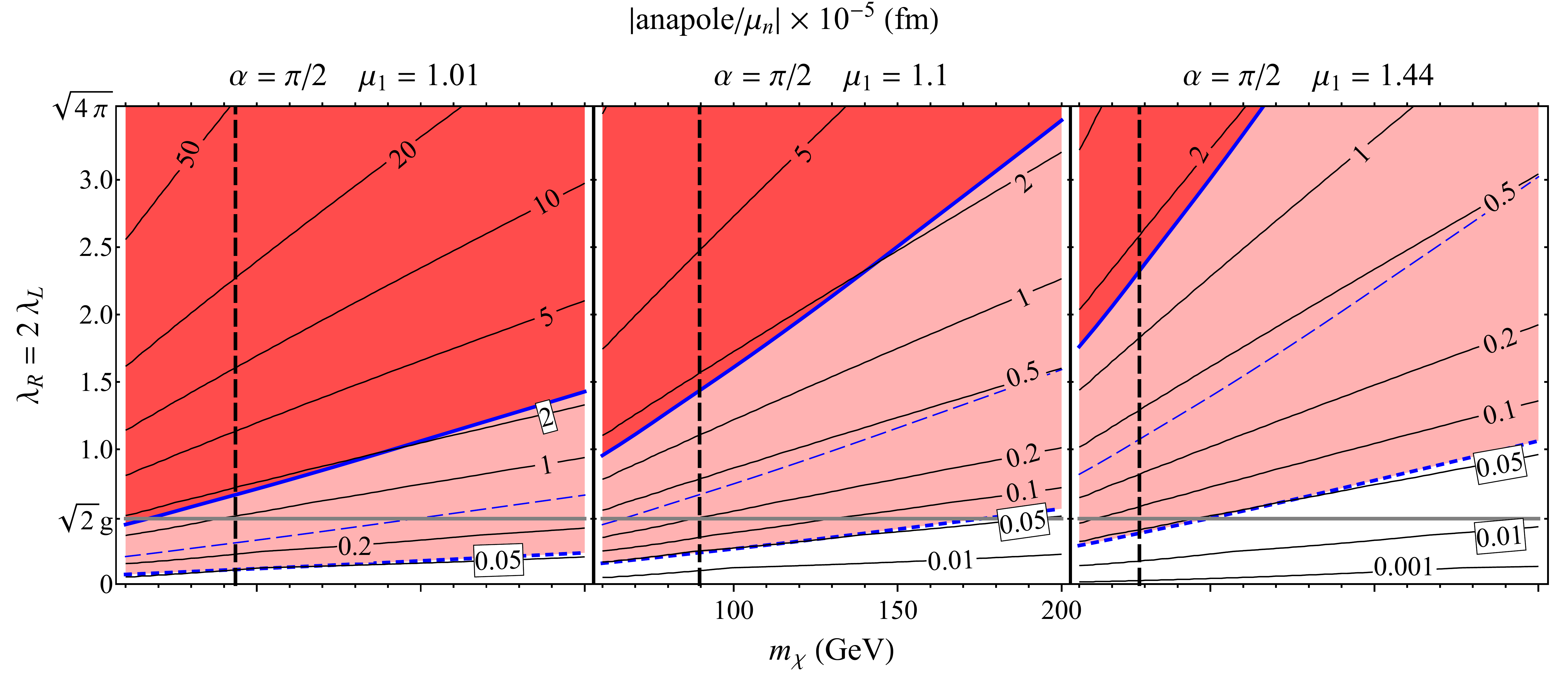}\\
  	\includegraphics[width=\textwidth]{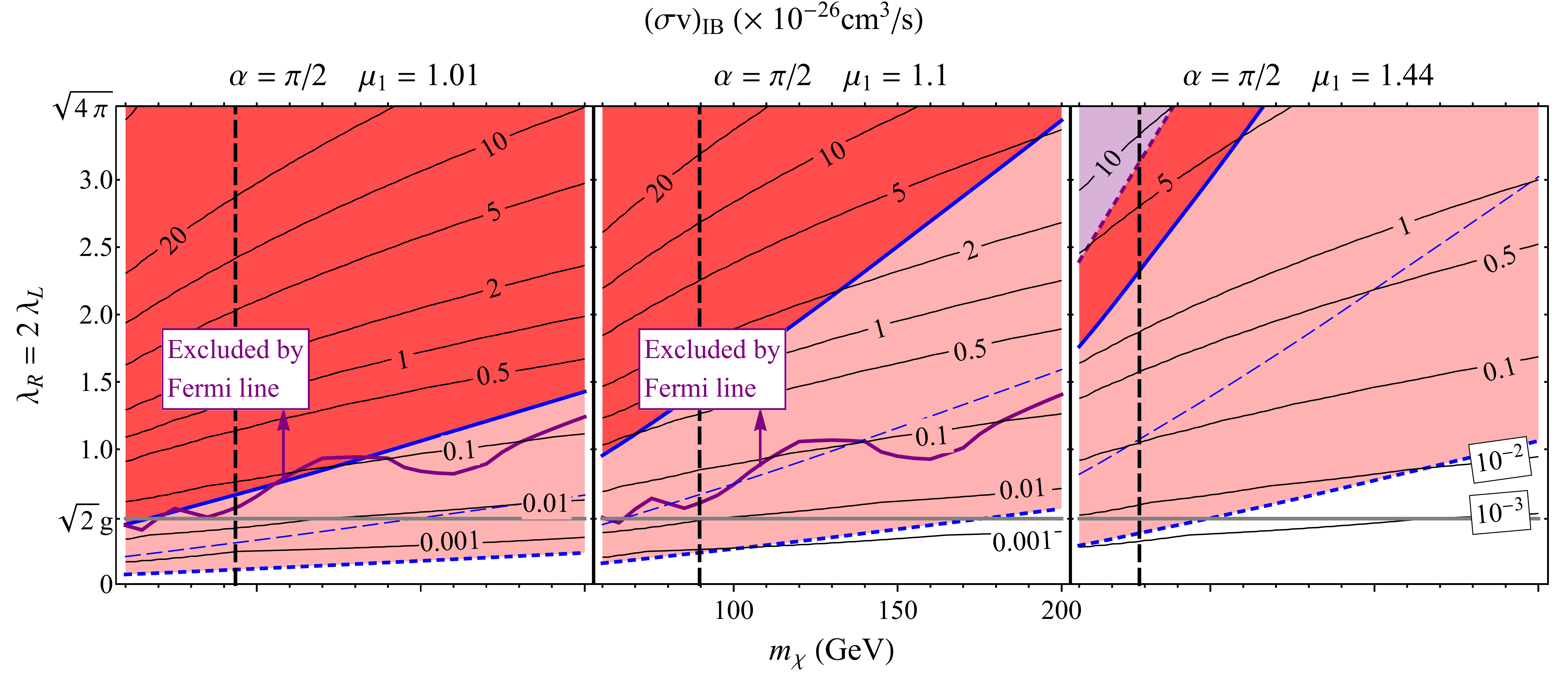}
\caption{$({m_\chi},{\lambda})$ for ${\alpha = \pi/2}$: We show plots of $\lambda_R = 2 \lambda_L$ versus $m_\chi$ for $\mu = 1.01$ (left), $\mu = 1.10$ (center), and $\mu = 1.44$ (right). In the first row, the contours correspond to values of $|\mathcal{A}/\mu_N| \times 10^{-5}$ fm. The vertical black dashed line is the LEP limit on the mass of the charged scalar for the $\mu$ channel.  The solid (dashed) blue lines correspond to LUX 2014 (future LZ) limits on the DM SI scattering cross section.  In addition, the most recent LUX 2016 constraint is estimated as a thin dashed blue contour.  The gray horizontal line corresponds to the SUSY value of couplings. In the second row, the red shaded regions and blue contours for LUX sensitivity remain the same, while additional black contours correspond to values of the IB cross section $(\sigma v)_{\text{IB}} \times 10^{-26}$ cm$^3$s$^{-1}$. The solid purple contour corresponds to the central value of the limits placed by $Fermi$-LAT on the DM annihilation cross section coming from dwarf galaxies, while the dashed purple contours correspond to $95\%$ CL interval. The thick purple horizontal line segments at $\alpha = 0, \pi/2,$ and $\pi$ correspond to limits from $Fermi$-LAT line searches.}
\label{Ymassalpha0andpiover4c}
\end{figure}

Our results for the LUX/LZ and $Fermi$-LAT sensitivities to our models are presented in the $(m_\chi,\lambda)$ planes in Figs.~\ref{Ymassalpha0andpiover4}, \ref{Ymassalpha0andpiover4b}, and \ref{Ymassalpha0andpiover4c}.

In Figure \ref{Ymassalpha0andpiover4}, we show our results in the $(m_\chi, \lambda)$ plane for $\alpha = 0$. As in Figs.~\ref{Yalpha100} and~\ref{Yalpha200}, $\lambda_R = 2\lambda_L$ and we show $\mu = 1.01$ (left), $\mu = 1.10$ (center), and $\mu = 1.44$ (right). Similarly, in the first row, the contours correspond to values of $|\mathcal{A}/\mu_N| \times 10^{-5}$ fm. The solid (dashed) blue lines correspond to LUX (future LZ) limits on the DM SI scattering cross section, with the estimation of the LUX 2016 limit represented as a thin dashed blue contour. The grey horizontal line corresponds to the SUSY value of couplings. In the second row, the contours correspond to values of the IB cross section $(\sigma v)_{\text{IB}} \times 10^{-26}$ cm$^3$s$^{-1}$, and the solid purple contour corresponds to limits from $Fermi$-LAT line searches. 

From the first row of Figure \ref{Ymassalpha0andpiover4}, we can see that a significant part of the parameter space up to DM mass $m_\chi = 200$ GeV is already being covered by LUX for $\mu = 1.01$, while LZ projections cover the parameter space almost entirely. The reaches degrade significantly for larger $\mu$, with LUX only covering a small part of the parameter space near large values of the Yukawa couplings. It is clear that LUX is unable to constrain the SUSY limit even in the best case scenario of $\mu = 1.01$. On the other hand, LZ covers almost the entire SUSY limit in this case, although the reach degrades for larger $\mu$. 

From the second row of Figure \ref{Ymassalpha0andpiover4}, we see that current line searches from $Fermi$-LAT are already sensitive to the same regions of parameter space that LUX is sensitive to for $\mu = 1.01$. For larger values of $\mu$, indirect detection considerably outperforms LUX, since, as before, the bounds from indirect detection are not strongly dependent on the degeneracy of the DM and mediator masses. LZ projections, however, continue to cover a larger parameter space than indirect detection.

 In Figure \ref{Ymassalpha0andpiover4b}, we repeat the results of Figure \ref{Ymassalpha0andpiover4} for $\alpha = \pi/4$.  We see that direct detection constraints are somewhat stronger than the $\alpha = 0$ case, in agreement with the first row of Figure \ref{Yalpha100}. In contrast to Figure \ref{Ymassalpha0andpiover4}, however, the appropriate indirect detection constraint to use for $\alpha = \pi/4$ are the $Fermi$-LAT continuum searches. We see that it outperforms LUX bounds for $\mu = 1.10$ and $\mu = 1.44$. LZ projections are stronger than indirect detection in both cases.

In Figure \ref{Ymassalpha0andpiover4c}, we show our results on the $(m_\chi, \lambda)$ plane for $\alpha = \pi/2$. From Figure \ref{Yalpha100}, it is clear that this value of $\alpha$ represents the best case scenario for direct detection. Indeed, we see that both LUX and LZ cover a substantially larger part of the parameter space as compared to the case of $\alpha = 0, \pi/4$ shown in previous figures. For $\mu = 1.01$, almost the entire parameter space up to $m_\chi = 200$ GeV is covered by LZ. LZ covers the entire SUSY limit up to $m_\chi = 200$ in the case of $\mu = 1.10$, and up to $m_\chi = 100$ GeV for $\mu = 1.44$. From the second row of Figure \ref{Ymassalpha0andpiover4c}, we see that while indirect detection limits (here, the appropriate constraint is from $Fermi$-LAT line searches) are comparable to LUX bounds for small $\mu = 1.01$, they perform vastly better for larger values of $\mu$. However, LZ still beats indirect detection limits.

\subsection{\texorpdfstring{$(\lambda_L,\lambda_R)$}{(lambdaL,lambdaR)} Planes}

\begin{figure}[t]
  \centering
\includegraphics[width=\textwidth]{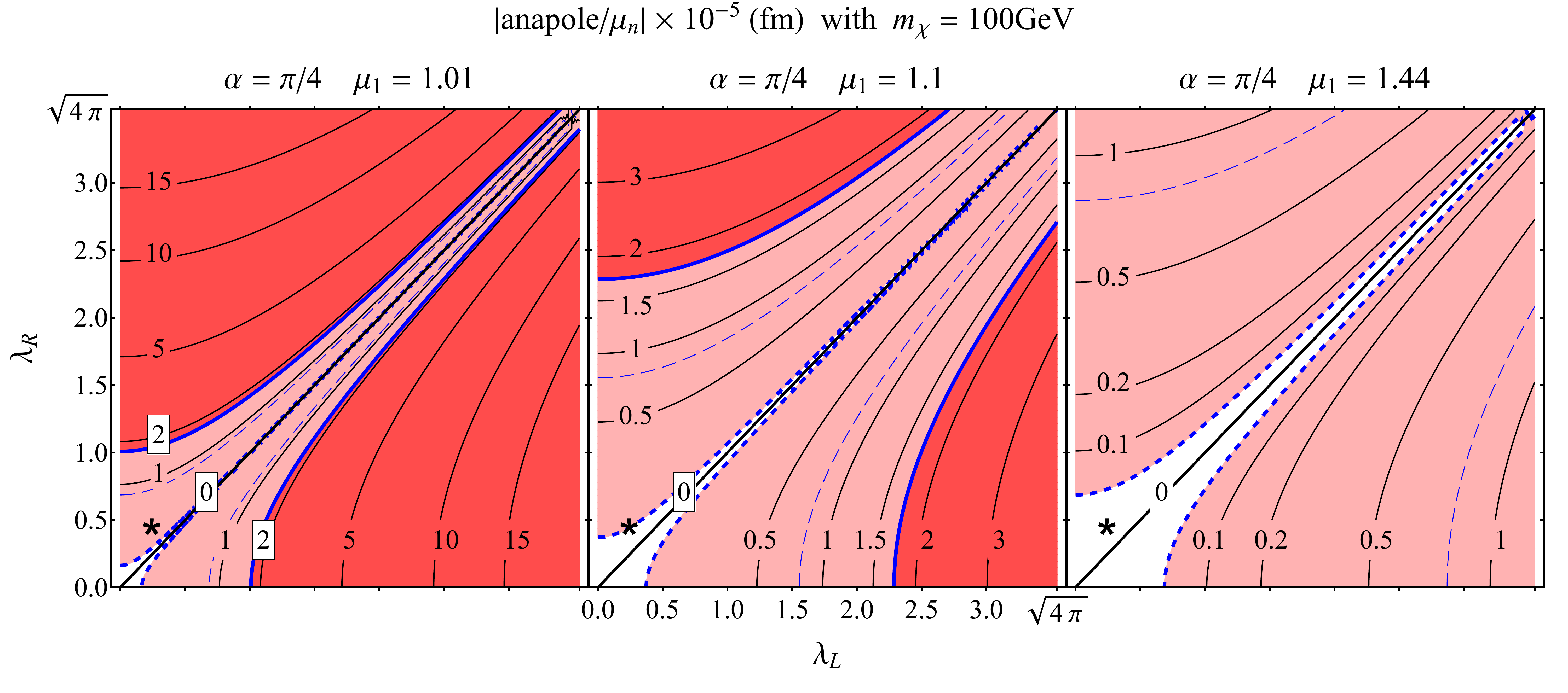}\\
\includegraphics[width=\textwidth]{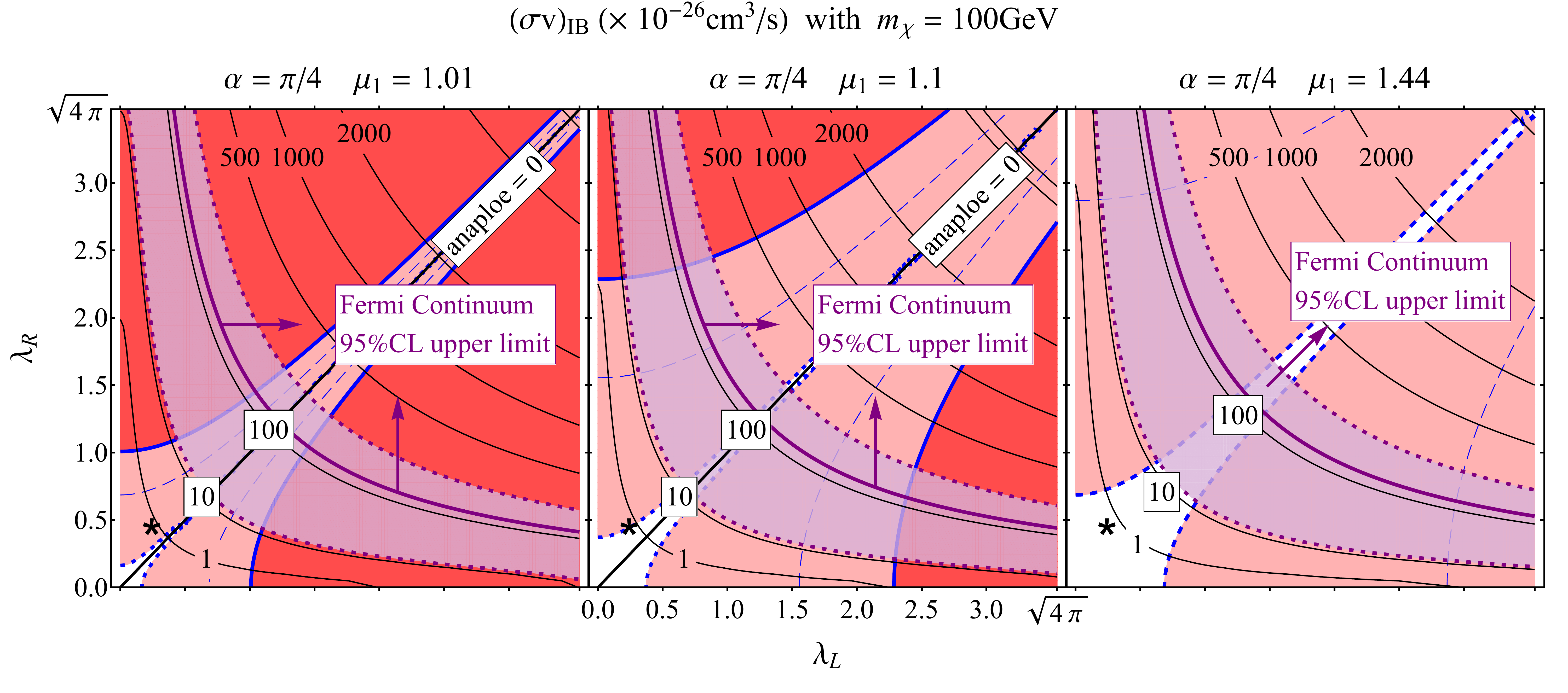}
  \caption{$({\lambda_L},{\lambda_R})$ for ${m_\chi = 100}$ GeV and ${\alpha=\pi/4}$: We show plots of $\lambda_R$ versus $\lambda_L$, for $\mu = 1.01$ (left), $\mu = 1.10$ (center), and $\mu = 1.44$ (right). In the first row, the contours correspond to values of $|\mathcal{A}/\mu_N| \times 10^{-5}$ fm. The solid (dashed) blue lines correspond to LUX 2014 (future LZ) limits on the DM SI scattering cross section.  In addition, the most recent LUX 2016 constraint is estimated as a thin dashed blue contour.  The gray horizontal line corresponds to the SUSY value of couplings. In the second row, the red shaded regions and blue contours for LUX sensitivity remain the same, while additional black contours correspond to values of the IB cross section $(\sigma v)_{\text{IB}} \times 10^{-26}$ cm$^3$s$^{-1}$. The solid purple contour corresponds to the central value of the limits placed by $Fermi$-LAT on the DM annihilation cross section coming from dwarf galaxies, while the dashed purple contours correspond to $95\%$ CL interval. The thick purple horizontal line segments at $\alpha = 0, \pi/2,$ and $\pi$ correspond to limits from $Fermi$-LAT line searches.}
  \label{ylyralphapiover4}
\end{figure}

\begin{figure}[t]
\centering
\includegraphics[width=\textwidth]{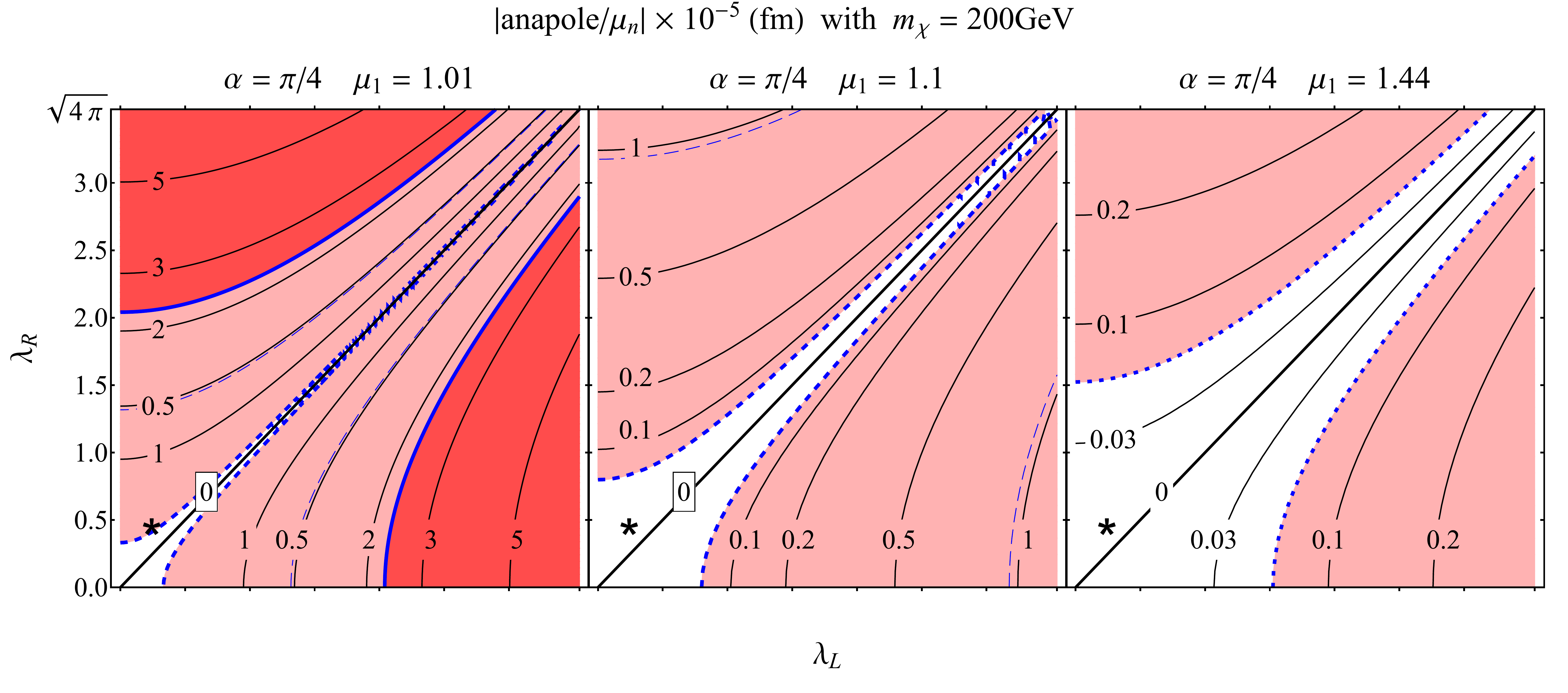}\\
\includegraphics[width=\textwidth]{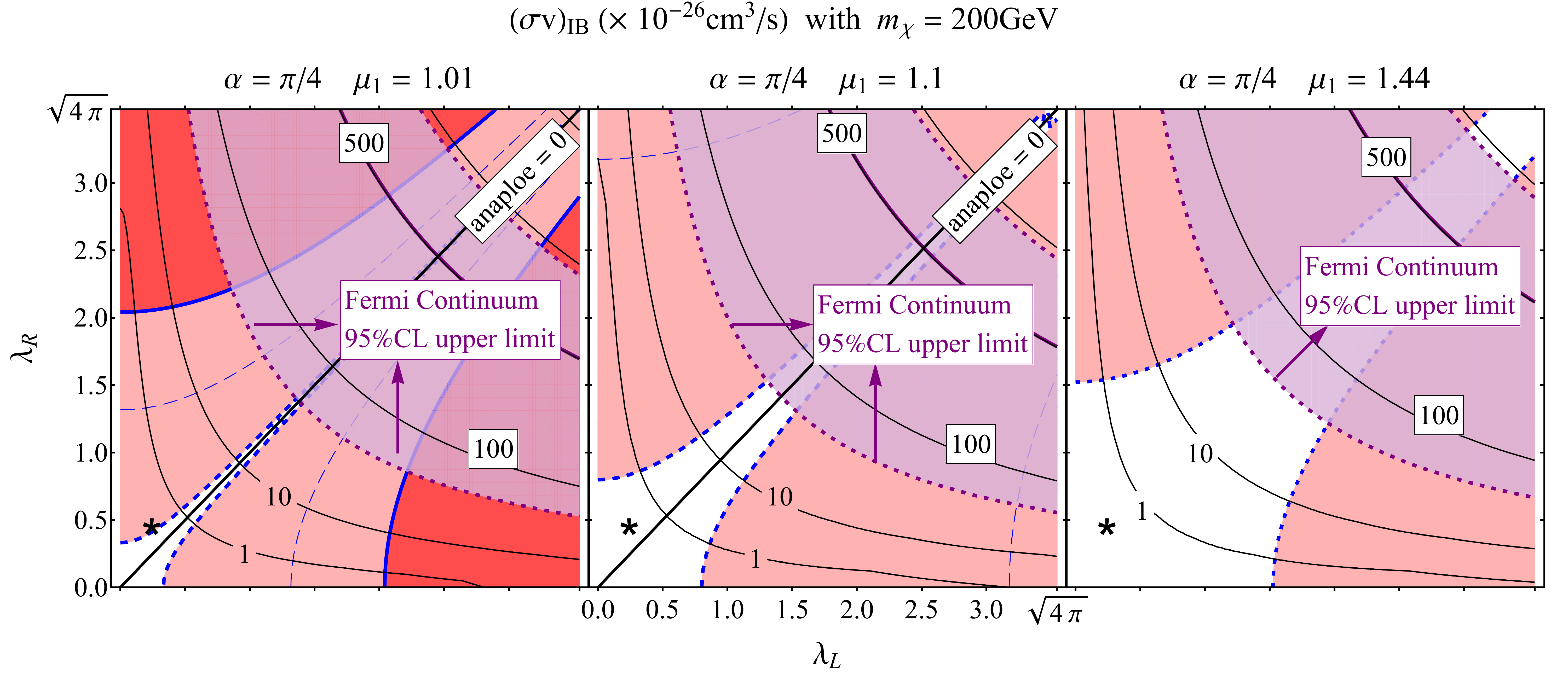}
\caption{$({\lambda_L},{\lambda_R})$ for ${m_\chi = 200}$ GeV and ${\alpha=\pi/4}$: We show plots of $\lambda_R$ versus $\lambda_L$, for $\mu = 1.01$ (left), $\mu = 1.10$ (center), and $\mu = 1.44$ (right). In the first row, the contours correspond to values of $|\mathcal{A}/\mu_N| \times 10^{-5}$ fm. The solid (dashed) blue lines correspond to LUX 2014 (future LZ) limits on the DM SI scattering cross section.  In addition, the most recent LUX 2016 constraint is estimated as a thin dashed blue contour.  The gray horizontal line corresponds to the SUSY value of couplings. In the second row, the red shaded regions and blue contours for LUX sensitivity remain the same, while additional black contours correspond to values of the IB cross section $(\sigma v)_{\text{IB}} \times 10^{-26}$ cm$^3$s$^{-1}$. The solid purple contour corresponds to the central value of the limits placed by $Fermi$-LAT on the DM annihilation cross section coming from dwarf galaxies, while the dashed purple contours correspond to $95\%$ CL interval. The thick purple horizontal line segments at $\alpha = 0, \pi/2,$ and $\pi$ correspond to limits from $Fermi$-LAT line searches.}
 \label{ylyralphapiover4b}
\end{figure}

\begin{figure}[t]
  \centering
  \includegraphics[width=\textwidth]{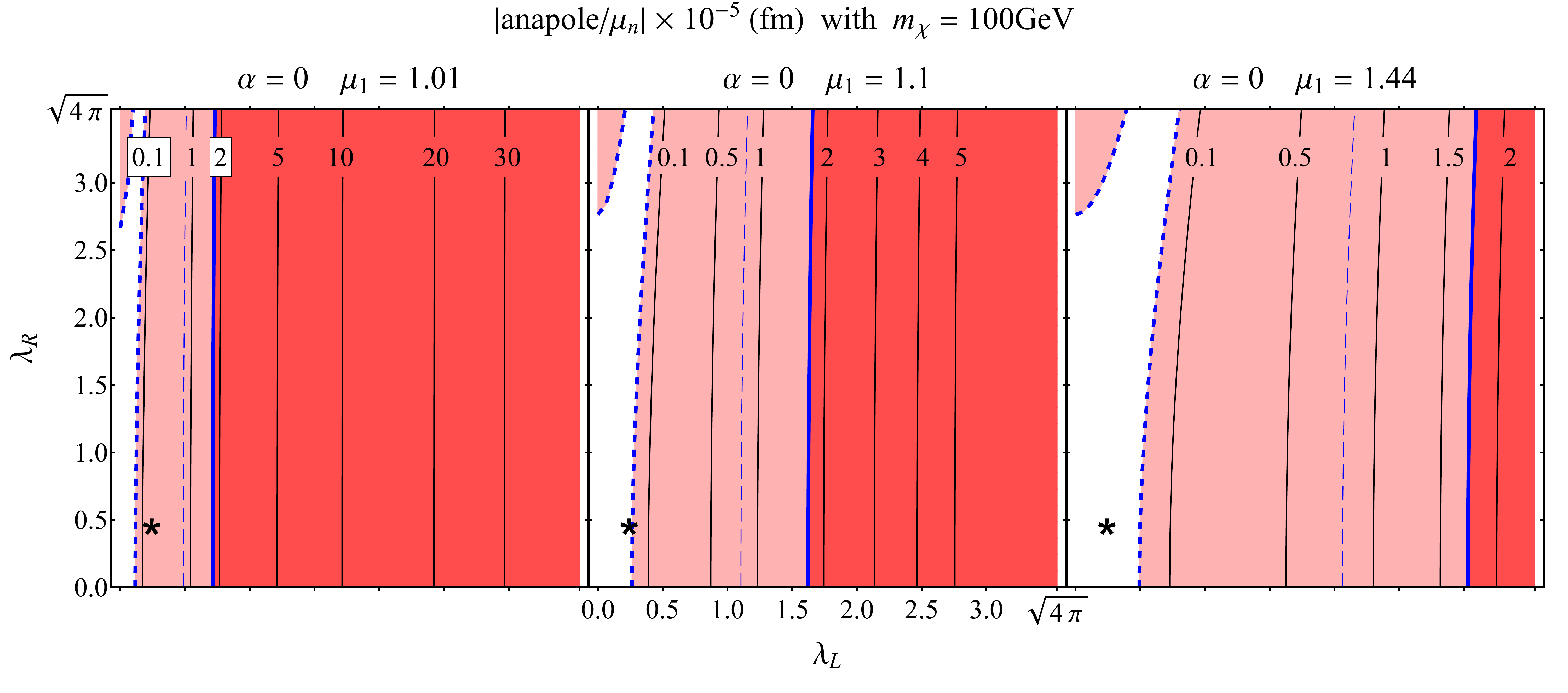}\\
  \includegraphics[width=\textwidth]{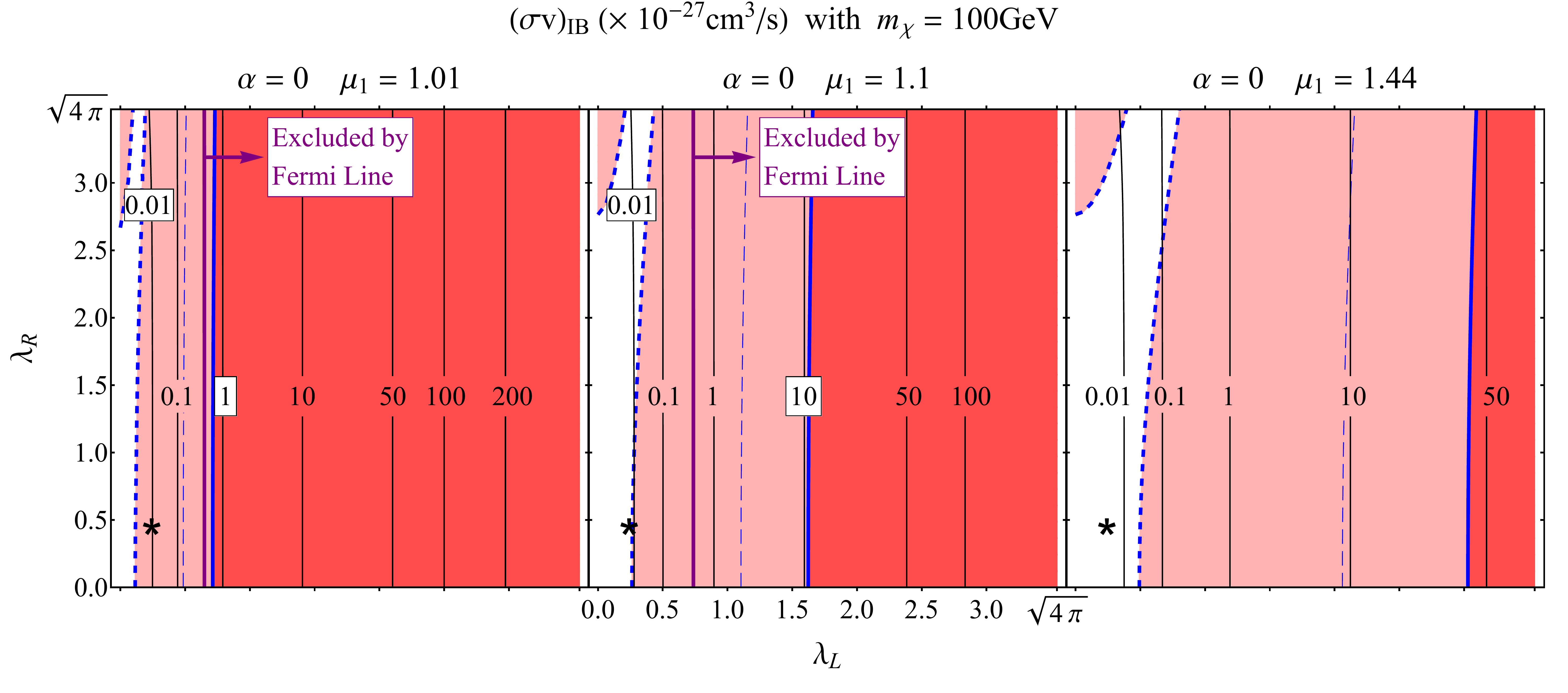}
  \caption{$({\lambda_L},{\lambda_R})$ for ${m_\chi = 100}$ GeV and ${\alpha=0}$: We show plots of $\lambda_R$ versus $\lambda_L$, for $\mu = 1.01$ (left), $\mu = 1.10$ (center), and $\mu = 1.44$ (right). In the first row, the contours correspond to values of $|\mathcal{A}/\mu_N| \times 10^{-5}$ fm. The solid (dashed) blue lines correspond to LUX 2014 (future LZ) limits on the DM SI scattering cross section.  In addition, the most recent LUX 2016 constraint is estimated as a thin dashed blue contour.  The gray horizontal line corresponds to the SUSY value of couplings. In the second row, the red shaded regions and blue contours for LUX sensitivity remain the same, while additional black contours correspond to values of the IB cross section $(\sigma v)_{\text{IB}} \times 10^{-26}$ cm$^3$s$^{-1}$. The solid purple contour corresponds to the central value of the limits placed by $Fermi$-LAT on the DM annihilation cross section coming from dwarf galaxies, while the dashed purple contours correspond to $95\%$ CL interval. The thick purple horizontal line segments at $\alpha = 0, \pi/2,$ and $\pi$ correspond to limits from $Fermi$-LAT line searches.}
  \label{ylyralpha0}
\end{figure}

\begin{figure}[t]
\centering
\includegraphics[width=\textwidth]{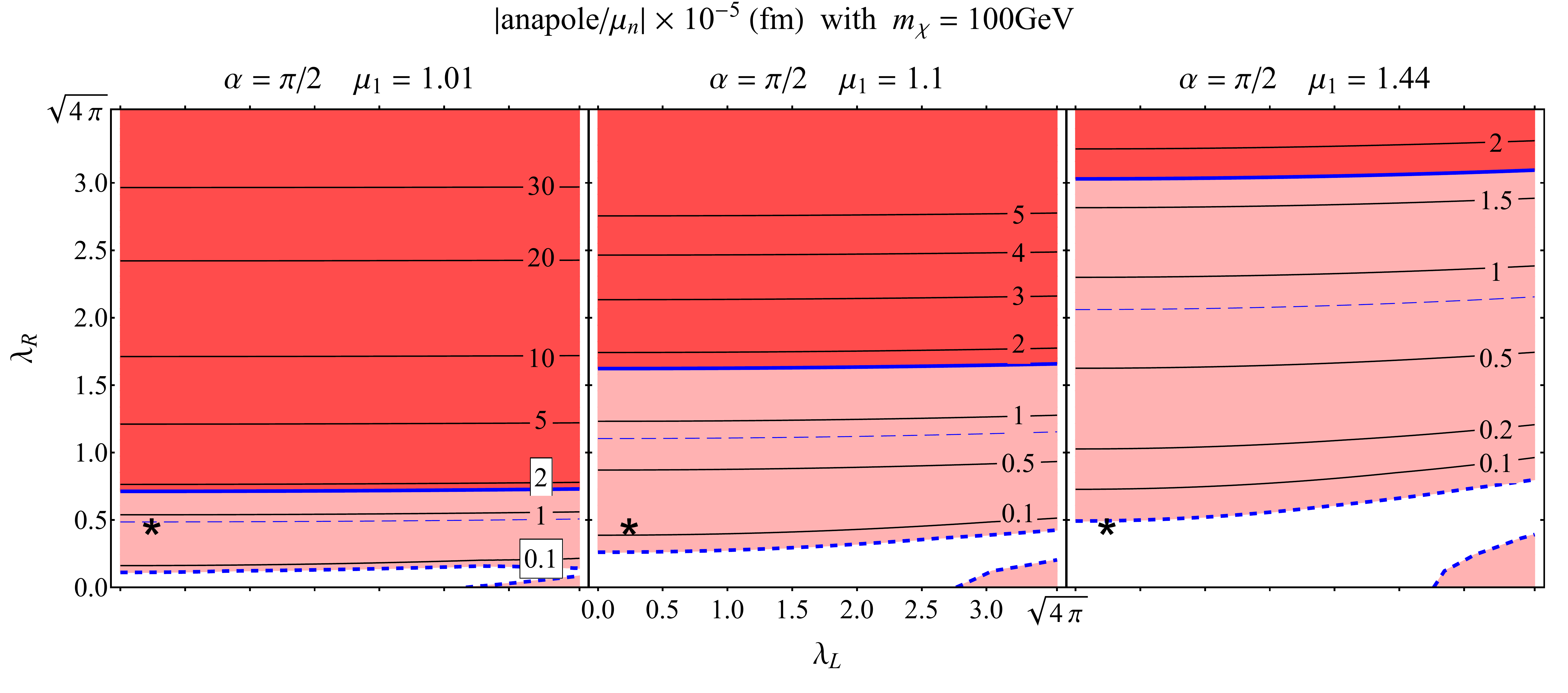}\\
\includegraphics[width=\textwidth]{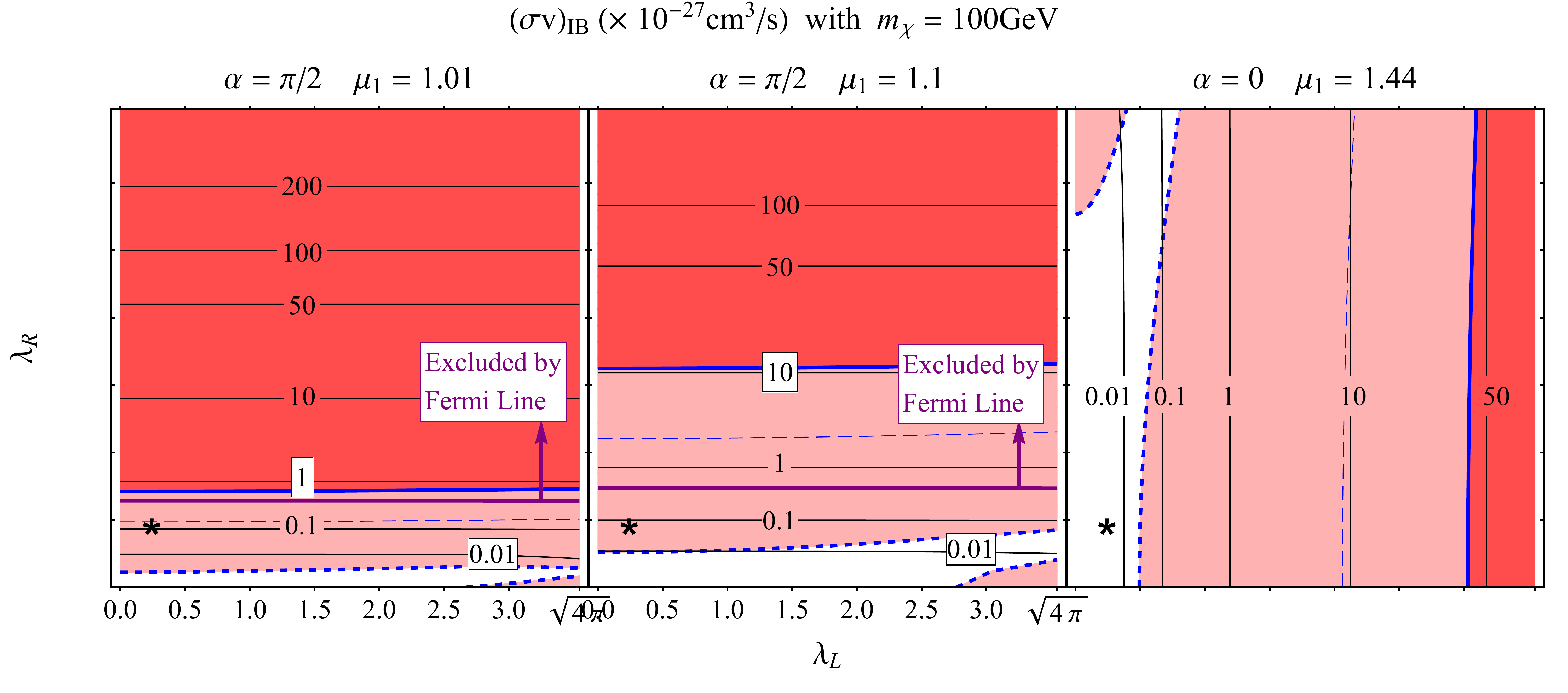}
\caption{$({\lambda_L},{\lambda_R})$ for ${m_\chi = 100}$ GeV and ${\alpha=\pi/2}$: We show plots of $\lambda_R$ versus $\lambda_L$, for $\mu = 1.01$ (left), $\mu = 1.10$ (center), and $\mu = 1.44$ (right). In the first row, the contours correspond to values of $|\mathcal{A}/\mu_N| \times 10^{-5}$ fm. The solid (dashed) blue lines correspond to LUX 2014 (future LZ) limits on the DM SI scattering cross section.  In addition, the most recent LUX 2016 constraint is estimated as a thin dashed blue contour.  The gray horizontal line corresponds to the SUSY value of couplings. In the second row, the red shaded regions and blue contours for LUX sensitivity remain the same, while additional black contours correspond to values of the IB cross section $(\sigma v)_{\text{IB}} \times 10^{-26}$ cm$^3$s$^{-1}$. The solid purple contour corresponds to the central value of the limits placed by $Fermi$-LAT on the DM annihilation cross section coming from dwarf galaxies, while the dashed purple contours correspond to $95\%$ CL interval. The thick purple horizontal line segments at $\alpha = 0, \pi/2,$ and $\pi$ correspond to limits from $Fermi$-LAT line searches.}
  \label{ylyralpha0b}
\end{figure}

Our results for the LUX/LZ and $Fermi$-LAT sensitivities to our models are presented in the $(\lambda_L,\lambda_R)$ planes in Figs.~\ref{ylyralphapiover4}, \ref{ylyralphapiover4b}, \ref{ylyralpha0}, and \ref{ylyralpha0b}.

In Figure \ref{ylyralphapiover4}, we display our plots on the $\lambda_R$ versus $\lambda_L$ plane keeping $m_\chi = 100$ GeV and $\alpha = \pi/4$ fixed, for $\mu = 1.01$ (left), $\mu = 1.10$ (center), and $\mu = 1.44$ (right). The star symbol corresponds to the SUSY value of couplings.  We see that most of the parameter space of the Yukawa couplings of our model is covered by a combination of direct and indirect detection. In fact, only the narrow corridor near $\lambda_R \sim \lambda_L$ constitutes a ``blind spot" where the anapole moment diminishes in magnitude for this particular value of $\alpha$. Exactly at $\lambda_R = \lambda_L$ the anapole moment vanishes and there are no direct detection constraints. The SUSY point is probed for $\mu = 1.10$ and below. For $\mu = 1.44$, LUX 2014 does not cover any part of the parameter space, while LZ covers most of it. From the second row, it is clear that indirect detection is able to probe a large portion of the corridor currently. In fact, parts of the parameter space where both $\lambda_R$ and $\lambda_L$ are larger than one are ruled out by $Fermi$-LAT. For $\mu = 1.44$, indirect detection is the only current bound on the parameter space. A combination of $Fermi$-LAT and LZ will rule out most of the plane even at $\mu = 1.44$, although the SUSY point will be beyond detection.

Figure \ref{ylyralphapiover4b} repeats the results of Figure \ref{ylyralphapiover4} for $m_\chi = 200$ GeV (again, $\alpha = \pi/4$). We see that both direct and indirect detection constraints become weaker, as expected. The LUX bounds are considerably weaker, and vanish entirely for $\mu = 1.10$. The corridor near the blind spot $\lambda_R = \lambda_L$ along which direct detection constraints are weak also gets wider. We see that indirect detection plays an important role in constraining the model for $m_\chi = 200$ GeV. Significant parts of the parameter space near the blind corridors are constrained by $Fermi$-LAT. 

In Figure \ref{ylyralpha0} and \ref{ylyralpha0b}, the results of Figure \ref{ylyralphapiover4} are repeated for the case of $\alpha = 0$ and  $\alpha = \pi/2$, respectively, in each case with $m_\chi = 100$ GeV.  We see that the corridor along which the scattering cross section is small has changed positions compared to the $\alpha = \pi/4$ case. For $\alpha = 0$ $(\pi/2)$, the corridor lies along $\lambda_L = 0 \, (\lambda_R = 0)$. For the $\alpha = 0$ case,  $\lambda_L \gtrsim 0.6$ $(1.6,\,3.0)$ is ruled out by LUX constraints for $\mu = 1.01\, (1.10,\, 1.44)$. On the other hand, $\lambda_L \gtrsim 0.1$ $(0.4,\,0.5)$ is ruled out by LZ constraints for $\mu = 1.01\, (1.10,\, 1.44)$. From the second row, it is clear that line search constraints from $Fermi$-LAT rule out the model for $\lambda_L \gtrsim 0.6$ for different values of $\mu$. This is comparable to the LUX limits for $\mu = 1.01$, but is better for larger values of $\mu$. It is clear, however, that the reach of LZ is better than that of indirect detection. The SUSY limit is constrained by LZ for $\mu = 1.01$. For $\mu > 1.10$, LZ  cannot constrain the SUSY limit.

\subsection{\label{sec:g-2}Muon \texorpdfstring{$g-2$}{g-2} constraints}

In our simplified model, the leading order contribution to the anomalous magnetic dipole moment $a=\frac{g-2}{2}$ is~\cite{Cheung:2009fc}, \cite{Bringmann:2012vr}:
\begin{equation}
  \Delta a_{f}=\frac{m_{f}m_{\chi}}{8\pi^{2}m^{2}_{\widetilde{f}_{1}}}\left|\lambda_{L}\lambda_{R}\right|\cos\varphi\cos\alpha\sin\alpha\left[\frac{1}{2(1-r_{1})^{2}}\left(1+r_{1}+\frac{2r_{1}\log r_{1}}{1-r_{1}}\right)\right]-(\widetilde{f}_{1}\rightarrow\widetilde{f}_{2})\,,
\end{equation}
where $r_{1}=m_{\chi}^{2}/m_{\widetilde{f}_{1}}^{2}$. In the $\mu$ channel, our simplified model will fully account for the $2\sigma$ deviation from the Standard Model~\cite{Hanneke:2008tm,Baron:2013eja,Bennett:2006fi,Bennett:2008dy,Abdallah:2003xd,Inami:2002ah}:
\begin{equation*}
  128\times 10^{-11}<\Delta a_{\mu}<448\times 10^{-11}
\end{equation*}
if \begin{enumerate*}[label=(\arabic*)]
\item $\varphi\sim\pi/2$ with arbitrary mixing angle $\alpha$, or 
\item $\alpha\sim 0$ or $\pi/2$ with arbitrary $\varphi$,
\end{enumerate*}
absent of fine-tuning in $\lambda_{L,R}$~\cite{Fukushima:2014yia,Kumar:2016cum}. For the $\tau$ channel, current experiments cannot put any sensitive limits on our parameter space. We note that the anomalous magnetic moment favors certain regions in our parameter space, but it does not put hard constraints on it. While direct detection experiments constrain the interaction of our simplified model with the SM, there are independent parameters in the new physics sector that tune $\Delta a_{\mu}$ to the observed value.

\section{\label{sec:con}Conclusion}

We have investigated simplified DM models coupled to SM fermions via charged mediators. We have considered the general case where fermionic DM couples to both right- and left-handed SM fermions, through two scalar mediators with arbitrary mixing angle $\alpha$. Results from direct detection for this class of models have been presented, and contrasted with results from indirect detection.

We note that the most stringent collider constraints for charged uncolored scalar particles come from LEP, and our study has been conducted for a spectrum which is beyond LEP bounds. The DM-nucleus scattering cross section in this class of models is mediated by higher electromagnetic moments, which, for Majorana DM, is the anapole moment. We give a full analytic derivation of the anapole moment for arbitrary $\alpha$ and match with limits presented in the literature. We then compute the scattering cross section, and translate bounds from LUX and LZ to the parameter space of the model. 

On the indirect detection side, we have presented the constraints coming from the $Fermi$-LAT line and continuum searches, after a careful discussion of  the chiral suppression of the annihilation cross-section in this class of models, and how it is lifted through either IB processes or non-zero mixing of the two mediators.

We have presented results for direct and indirect detection and found that they probe complementary regions of parameter space. Results in the supersymmetric limit of these simplified models are provided in all cases. We have found that future direct detection experiments like LZ will probe a significant portion of the parameter space of these models for $m_\chi \sim \mathcal{O}(100-200)$ GeV and lightest mediator mass within $\mathcal{O}(5\%)$ of the DM mass. However, the direct detection prospects become weaker for larger DM mass and larger mass gap between the DM and the lightest mediator mass. At DM mass of $\mathcal{O}(200)$ GeV and lightest mediator mass $\sim 20\%$ larger than the DM mass, direct detection constraints are already too feeble to probe the SUSY limit of these models. The direct detection bounds also disappear at certain ``blind spots" in the parameter space, where the anapole moment vanishes or nearly vanishes. These regions have been carefully studied.

Generally, we have found that current $Fermi$-LAT and LUX 2014 results have comparable reaches in this class of models, for $\mu = 1.01$. However, for larger $\mu$, the indirect detection constraints start to become more constraining than LUX. Indirect detection is also able to constrain regions of parameter space where the blind spots occur. Future LZ projections generally outperform indirect detection constraints for all choices of parameters, except at the blind spots.

It is interesting to contrast our work with that of models with simpler mediator sectors, such as the case of a single scalar mediator coupling to right-handed SM fermions considered in \cite{Kopp:2014tsa}. This corresponds to a choice of $\alpha = \pi/2$ in the models presented here. While the dependence of the anapole moment on the mixing angle is quite simple, there are several new physical features that emerge when one considers the more elaborate mediator sector. These features are evident in Figure 2 and have been discussed throughout the paper. For example, it is clear that the case of a single mediator coupling to right-handed fermions ($\alpha \rightarrow \pi/2$) actually affords the most optimistic outlook in terms of direct detection. The prospects dwindle rapidly as $\alpha$ is changed, until one reaches the blind spots where they are very weak and one must rely on indirect detection to constrain the model. This more general mediator sector thus displays the complementarity of direct and indirect detection, which is one of the main themes of the paper.

\section{Acknowledgements}

We would like to thank Takahiro Yamamoto for collaboration in the early stages of this work. We would also like to thank Paolo Gondolo, Jason Kumar and Juri Smirnov for helpful discussions, and Cora Kaiser for cooperation and encouragement.  PS is supported in part by NSF Grant No. PHY-1417367.

\appendix

\section{\label{sec:fullana}Full Analytic Expression of the Anapole Moment}
In this appendix, we derive the full analytic expression of the DM anapole moment from the Lagrangian \eqref{eq:Lint} and the relevant QED interaction,
\begin{equation}
\mathcal{L}_{\text{qed}}=ie\left(\widetilde{f}_{1}^{\ast}A^{\mu}\partial_{\mu}\widetilde{f}^{}_{1}+\widetilde{f}_{2}^{\ast}A^{\mu}\partial_{\mu}\widetilde{f}^{}_{2}-\text{c.c}\right)+e\,\overline{f}\gamma^{\mu}A_{\mu}f.
\end{equation}
The total off-shell amplitude $\mathcal{M}^{\mu}$ is the sum of all four Feynman diagrams in Fig.~\ref{fig:FeynAnapole},
\begin{equation}
  \mathcal{M}^{\mu}=\mathcal{M}^{\mu}_{1}+\mathcal{M}^{\mu}_{2}+\mathcal{M}^{\mu}_{3}+\mathcal{M}^{\mu}_{4}.
\end{equation}
Each $\mathcal{M}_{i}^{\mu}$ contains the contribution from two internal scalars $\widetilde{f}_{1}$ and $\widetilde{f}_{2}$, namely,
\begin{equation*}
  \mathcal{M}^{\mu}_{k}=\mathcal{M}^{\mu}_{k}(1)+\mathcal{M}^{\mu}_{k}(2).
\end{equation*}
The Majorana nature of $\chi$ requires that $\mathcal{M}^{\mu}$ must have the form of Eq.~\eqref{eq:Mmu}, and we are going to explicitly show this. We only need to calculate $\mathcal{M}^{\mu}_{k}(1)$, from which $\mathcal{M}^{\mu}_{k}(2)$ can be obtained through the replacement
\begin{align*}
  &m_{\widetilde{f}_{1}}\rightarrow m_{\widetilde{f}_{2}}& &\cos\alpha\rightarrow\sin\alpha& &\sin\alpha\rightarrow -\cos\alpha.
\end{align*}
If we call the undetermined loop momentum $k$ in all these diagrams, the sub-amplitudes can be expressed as
\begin{align}
  &\mathcal{M}^{\mu}_{1}(1)=\int\frac{d^{4}k}{(2\pi)^{4}}\frac{\mathcal{F}_{1}^{\mu}+m_{f}\mathcal{G}^{\mu}}{d_{1}(1)d_{2}(1)d_{3}(1)}& &\mathcal{M}^{\mu}_{2}(1)=\int\frac{d^{4}k}{(2\pi)^{4}}\frac{\mathcal{F}_{2}^{\mu}-m_{f}\mathcal{G}^{\mu}}{d_{1}(1)d_{2}(1)d_{3}(1)}\nonumber\\
  &\mathcal{M}^{\mu}_{3}(1)=\int\frac{d^{4}k}{(2\pi)^{4}}\frac{\mathcal{F}_{3}^{\mu}+m_{f}\mathcal{H}^{\mu}}{\widetilde{d}_{1}(1)\widetilde{d}_{2}(1)\widetilde{d}_{3}(1)}& &\mathcal{M}^{\mu}_{4}(1)=\int\frac{d^{4}k}{(2\pi)^{4}}\frac{\mathcal{F}_{4}^{\mu}-m_{f}\mathcal{H}^{\mu}}{\widetilde{d}_{1}(1)\widetilde{d}_{2}(1)\widetilde{d}_{3}(1)}.
\end{align}
In these equations, $d$'s are the propagator denominators,
\begin{align*}
  &d_{1}(1)=k^{2}-m^{2}_{\widetilde{f}_{1}}& &d_{2}(1)=(k+p)^{2}-m^{2}_{f}& &d_{3}(1)=(k+p')^{2}-m^{2}_{f},
\end{align*}
and $\widetilde{d}$ are obtained by exchanging the scalar mass $m_{\widetilde{f}_{1}}$ and the fermion mass $m_{f}$. In the numerators, the fermion chains are
\begin{align}
  &\mathcal{F}_{1}^{\mu}=-\overline{u}(p')\left(|\lambda_{L}|^{2}\cos^{2}\alpha\,P_{L}+|\lambda_{R}|^{2}\sin^{2}\alpha\,P_{R}\right)[(\slashed{k}+\slashed{p}')\gamma^{\mu}(\slashed{k}+\slashed{p})+m^{2}_{f}\gamma^{\mu}]u(p)\nonumber\\
  &\mathcal{F}_{2}^{\mu}=\overline{u}(p')\left(|\lambda_{L}^{2}|\cos^{2}\alpha\,P_{R}+|\lambda_{R}|^{2}\sin^{2}\alpha\,P_{L}\right)[(\slashed{k}+\slashed{p}')\gamma^{\mu}(\slashed{k}+\slashed{p})+m^{2}_{f}\gamma^{\mu}]u(p)\nonumber\\
  &\mathcal{F}_{3}^{\mu}=-(2k+p+p')^{\mu}\overline{u}(p')\left(|\lambda_{L}|^{2}\cos^{2}\alpha\,P_{L}+|\lambda_{R}|^{2}\sin^{2}\alpha\,P_{R}\right)\slashed{k}\,u(p)\nonumber\\
  &\mathcal{F}_{4}^{\mu}=(2k+p+p')^{\mu}\overline{u}(p')\left(|\lambda_{L}|^{2}\cos^{2}\alpha\,P_{R}+|\lambda_{R}|^{2}\sin^{2}\alpha\,P_{L}\right)\slashed{k}\,u(p)
\end{align}
\begin{align}
  &\mathcal{G}^{\mu}=-|\lambda_{L}\lambda_{R}|\sin\alpha\cos\alpha\,\overline{u}(p')(e^{i\varphi}P_{L}+e^{-i\varphi}P_{R})[(\slashed{k}+\slashed{p}')\gamma^{\mu}+\gamma^{\mu}(\slashed{k}+\slashed{p})]u(p)\nonumber\\
  &\mathcal{H}^{\mu}=|\lambda_{L}\lambda_{R}|\sin\alpha\cos\alpha\,(2k+p+p')^{\mu}\overline{u}(p')(e^{i\varphi}P_{L}+e^{-i\varphi}P_{R})u(p).
\end{align}
In the total amplitude $\mathcal{M}^{\mu}$, the $\mathcal{G}^{\mu}$ and $\mathcal{H}^{\mu}$ parts cancel, leaving
\begin{equation}
  \mathcal{M}^{\mu}=e\left(|\lambda_{L}|^{2}\cos^{2}\alpha-|\lambda_{R}|^{2}\sin^{2}\alpha\right)\mathcal{I}_{1}^{\mu}+e\left(|\lambda_{L}|^{2}\sin^{2}\alpha-|\lambda_{R}|^{2}\cos^{2}\alpha\right)\mathcal{I}_{2}^{\mu},
\end{equation}
where $\mathcal{I}^{\mu}_{i}$ is
\begin{align}
  \mathcal{I}^{\mu}_{i}=\int\frac{d^{4}k}{(2\pi)^{4}}\left[\frac{(2k+p+p')^{\mu}\overline{u}(p')\gamma^{5}\slashed{k}\,u(p)}{d_{1}(i)d_{2}(i)d_{3}(i)}+\frac{\overline{u}(p')\gamma^{5}[(\slashed{k}+\slashed{p}')\gamma^{\mu}(\slashed{k}+\slashed{p})+m^{2}_{f}\gamma^{\mu}]u(p)}{\widetilde{d}_{1}(i)\widetilde{d}_{2}(i)\widetilde{d}_{3}(i)}\right].
\end{align}
{The cancelation of $\mathcal{G}^{\mu}$ and $\mathcal{H}^{\mu}$ can be understood in the following way. As we have noted before Eq.~\eqref{eq:analag}, $\mathcal{A}$ should be real in nature so that $\mathcal{G}^{\mu}$ and $\mathcal{H}^{\mu}$, containing the factor $e^{i\varphi}P_{L}+e^{-i\varphi}P_{R}$ that introduce an imaginary part $i\sin\varphi$, must cancel by themselves.}

Using some spinor and $\gamma$-matrix identities, we can rewrite $\mathcal{I}^{\mu}$ into
\begin{align}
  \mathcal{I}^{\mu}_{i}=i\,\overline{u}(p')(Y_{i}\gamma^{\mu}-X_{i}\slashed{q}q^{\mu})\gamma^{5}u(p),
\end{align}
where $X_{i}$ and $Y_{i}$ can be expanded with respect to tensor loop integrals,
\begin{align}
  X_{i}&=\mathcal{C}_{11}(i)+\mathcal{C}_{1}(i)+\widetilde{\mathcal{C}}_{11}(i)-\mathcal{C}_{12}(i)-\widetilde{\mathcal{C}}_{12}(i)\nonumber\\
  Y_{i}&=2\mathcal{C}_{00}(i)-2\widetilde{\mathcal{C}}_{00}(i)+2m_{\chi}^{2}\mathcal{C}_{11}(i)+(2m_{\chi}^{2}-q^{2})\mathcal{C}_{12}(i)\nonumber\\
  &\quad+4m_{\chi}^{2}\mathcal{C}_{1}(i)+(m_{\chi}^{2}-m_{f}^{2})\mathcal{C}_{0}(i).
\end{align}
The loop integrals $\mathcal{C}$ and $\widetilde{\mathcal{C}}$ are related to those $3$-point integrals $C$ defined in \texttt{LoopTools} \cite{Hahn:1998yk} through
\begin{align}
  &\mathcal{C}_{(\cdots)}(i)\equiv\frac{1}{16\pi^{2}}\,C_{(\cdots)}[\,m^{2}_{\chi},q^{2},m^{2}_{\chi},m_{\widetilde{f}_{i}}^{2},m_{f}^{2},m_{f}^{2}\,]\nonumber\\
  &\widetilde{\mathcal{C}}_{(\cdots)}(i)\equiv\frac{1}{16\pi^{2}}\,C_{(\cdots)}[\,m^{2}_{\chi},q^{2},m^{2}_{\chi},m^{2}_{f},m^{2}_{\widetilde{f}_{i}},m^{2}_{\widetilde{f}_{i}}\,].
\end{align}
Then, using the techniques reviewed in \cite{Ellis:2011cr}, we can expand the tensor and vector loop integrals in terms of the scalar ones ($\mathcal{C}_{0}$ and $\widetilde{\mathcal{C}}_{0}$) and $2$-point integrals. The result is that
\begin{align}
\label{eq:X_i}
  -\xi^{2}_{\chi}X_{i}&=(1-\delta)\mathcal{C}_{0}(i)+(1-\mu_{i})\widetilde{\mathcal{C}}_{0}(i)+(3-\mu_{i}+\delta)\mathcal{C}_{1}(i)\nonumber\\
  &\quad+(\xi^{2}_{\chi}+\mu_{i}-\delta+3)\widetilde{\mathcal{C}}_{1}(i)+\mathcal{B}_{0}(i)\,,
\end{align}
where $\xi_{\chi}^{2}=-q^{2}/m_{\chi}^{2}$. In this process, we may also prove that $Y_{i}=q^{2}X_{i}$. As a result, we can arrive at
\begin{equation}
  \mathcal{I}^{\mu}_{i}=iX_{i}\overline{u}(p')(q^{2}\gamma^{\mu}-\slashed{q}q^{\mu})\gamma^{5}u(p),
\end{equation}
and consequently Eq.~\eqref{eq:Mmu} and \eqref{eq:A_alpha}. In Eq.~\eqref{eq:X_i}, $\mathcal{B}_{0}$ is a combination of $2$-point loop integrals whose divergent parts cancel each other,
\begin{align}
  \mathcal{B}_{0}(i)=\frac{1}{2m^{2}_{\chi}}&\left[2-(\mu_{i}-\delta)\log\left(\frac{\mu_{i}}{\delta}\right)+2\sqrt{\Delta_{i}}\arctanh\left(\frac{\sqrt{\Delta_{i}}}{\mu_{i}+\delta-1}\right)\right.\nonumber\\
    &\left.+2\sqrt{\frac{4\delta+\xi_{\chi}^{2}}{\xi_{\chi}^{2}}}\arctanh\sqrt{\frac{\xi_{\chi}^{2}}{4\delta+\xi_{\chi}^{2}}}+2\sqrt{\frac{4\mu_{i}+\xi_{\chi}^{2}}{\xi_{\chi}^{2}}}\arctanh\sqrt{\frac{\xi_{\chi}^{2}}{4\mu_{i}+\xi_{\chi}^{2}}}\right].
\end{align}
The vector loop integrals $\mathcal{C}_{1}$ and $\widetilde{\mathcal{C}}_{1}$ can be written as a combination of $2$-point integrals and scalar $3$-point integrals,
\begin{align}
  &\mathcal{C}_{1}(i)=\left(\frac{\delta-\mu_{i}-1}{4+\xi_{\chi}^{2}}\right)\mathcal{C}_{0}+\mathcal{B}_{1}(i)& &\widetilde{\mathcal{C}}_{1}=\left(\frac{\mu_{i}-\delta-1}{4+\xi_{\chi}^{2}}\right)\widetilde{\mathcal{C}}_{0}+\widetilde{\mathcal{B}}_{1}(i),
\end{align}
where $\mathcal{B}_{1}$ and $\widetilde{\mathcal{B}}_{1}$ are
\begin{align}
  \mathcal{B}_{1}(i)=\frac{1}{m_{\chi}^{2}(4+\xi_{\chi}^{2})}&\left[-\frac{\mu_{i}-\delta+1}{2}\log\left(\frac{\mu_{i}}{\delta}\right)+\sqrt{\Delta_{i}}\arctanh\left(\frac{\sqrt{\Delta_{i}}}{\mu_{i}+\delta-1}\right)\right.\nonumber\\
    &\;\left.+\sqrt{\frac{\xi_{\chi}^{2}+4\delta}{\xi_{\chi}^{2}}}\arctanh\sqrt{\frac{\xi_{\chi}^{2}}{4\delta+\xi_{\chi}^{2}}}\,\right]
\end{align}
\begin{align}
  \widetilde{\mathcal{B}}_{1}(i)=\frac{1}{m_{\chi}^{2}(4+\xi_{\chi}^{2})}&\left[-\frac{\mu_{i}-\delta-1}{2}\log\left(\frac{\mu_{i}}{\delta}\right)+\sqrt{\Delta_{i}}\arctanh\left(\frac{\sqrt{\Delta_{i}}}{\mu_{i}+\delta-1}\right)\right.\nonumber\\
    &\left.+\sqrt{\frac{4\mu_{i}+\xi_{\chi}^{2}}{\xi_{\chi}^{2}}}\arctanh\sqrt{\frac{\xi_{\chi}^{2}}{4\mu_{i}+\xi_{\chi}^{2}}}\,\right].
\end{align}
Now we are left with the last two pieces of $X_{i}$, $\mathcal{C}_{0}$ and $\widetilde{\mathcal{C}}_{0}$. They can be calculated using the technique developed in \cite{'tHooft:1978xw}. To write them in a compact form, we introduce the following variables:
\begin{align}
  &x_{1,2}^{i}=-\frac{(\mu_{i}-\delta-1)\pm\sqrt{\Delta_{i}}}{2}& &\widetilde{x}_{1,2}^{i}=-\frac{(\delta-\mu_{i}-1)\pm\sqrt{\Delta_{i}}}{2}\nonumber\\
  &z_{1,2}^{i}=\frac{\xi_{\chi}\pm\sqrt{4\delta+\xi_{\chi}}}{2\xi_{\chi}}& &\widetilde{z}_{1,2}^{i}=\frac{\xi_{\chi}\pm\sqrt{4\mu_{i}+\xi_{\chi}}}{2\xi_{\chi}}.
\end{align}
In the above variables, we implicitly assign an infinitesimal imaginary part $-i\epsilon$ to those with subscript $1$ and $+i\epsilon$ to those with subscript $2$ when necessary. This imaginary part is important for analytic continuation beyond the branching points of the logarithm and dilogarithm functions. In addition, we have
\begin{align}
	& y_{1}^{i}=\frac{1-(1-a)(\mu_{i}-\delta)}{2-a}& & \widetilde{y}_{1}^{i}=\frac{1-(1-a)(\delta-\mu_{i})}{2-a}\nonumber\\
	& y_{2}^{i}=\frac{1-a-\mu_{i}+\delta}{2-a}& & \widetilde{y}_{2}^{i}=\frac{1-a-\delta+\mu_{i}}{2-a}\nonumber\\
	& y_{3}^{i}=\frac{a(\mu_{i}-\delta-1+a)}{(2-a)\xi_{\chi}^{2}}& & \widetilde{y}_{3}^{i}=\frac{a(\delta-\mu_{i}-1+a)}{(2-a)\xi_{\chi}^{2}},
\end{align}
where $a=\frac{-\xi_{\chi}^{2}+\sqrt{4+\xi_{\chi}^{2}}}{2}$. Effectively, the variables with a tilde are obtained by exchanging $\delta$ and $\mu_{i}$ in those without a tilde. These variables appear as arguments of dilogarithm functions in
\begin{align}
	& I^{i}_{1}= \Li_{2}\left(\frac{y^{i}_{1}}{y^{i}_{1}-x^{i}_{1}}\right)-\Li_{2}\left(\frac{y^{i}_{1}-1}{y^{i}_{1}-x^{i}_{1}}\right)+ \Li_{2}\left(\frac{y^{i}_{1}}{y^{i}_{1}-x^{i}_{2}}\right)-\Li_{2}\left(\frac{y^{i}_{1}-1}{y^{i}_{1}-x^{i}_{2}}\right) \nonumber\\
	& I^{i}_{2}= \Li_{2}\left(\frac{y^{i}_{2}}{y^{i}_{2}-x^{i}_{1}}\right)-\Li_{2}\left(\frac{y^{i}_{2}-1}{y^{i}_{2}-x^{i}_{1}}\right)+\Li_{2}\left(\frac{y^{i}_{2}}{y^{i}_{2}-x^{i}_{2}}\right)-\Li_{2}\left(\frac{y^{i}_{2}-1}{y^{i}_{2}-x^{i}_{2}}\right)\nonumber\\
	& I^{i}_{3}= \Li_{2}\left(\frac{y^{i}_{3}}{y^{i}_{3}-z^{i}_{1}}\right)-\Li_{2}\left(\frac{y^{i}_{3}-1}{y^{i}_{3}-z^{i}_{1}}\right)+\Li_{2}\left(\frac{y^{i}_{3}}{y^{i}_{3}-z^{i}_{2}}\right)-\Li_{2}\left(\frac{y^{i}_{3}-1}{y^{i}_{3}-z^{i}_{2}}\right).
\end{align}
We also have $\widetilde{I}_{1,2,3}^{i}$, in which the dilogarithm functions have $\widetilde{x}$, $\widetilde{y}$ and $\widetilde{z}$ as variables. Finally, in terms of $I_{1,2,3}$ and $\widetilde{I}_{1,2,3}$, we simply have
\begin{align}
	&\mathcal{C}_{0}(i)=-b\,(I_{1}^{i}-I_{2}^{i}+I_{3}^{i})& &\widetilde{\mathcal{C}}_{0}(i)=-b\,(\widetilde{I}^{i}_{1}-\widetilde{I}_{2}^{i}+\widetilde{I}_{3}^{i}),
\end{align}
where $b=\frac{1}{m_{\chi}^{2}\,\xi_{\chi}\sqrt{4+\xi_{\chi}^{2}}}$. In the limit $\xi_{\chi}\rightarrow 0$, it is tedious but still straightforward to verify that the leading term on the right hand side of Eq.~\eqref{eq:X_i} is $\mathcal{O}(\xi_{\chi}^{2})$ such that $X_{i}$ is independent of $q^{2}$ in this limit:
\begin{equation}
X_{i}\approx\frac{1}{96\pi^{2}m_{\chi}^{2}}\left[\frac{3\mu_{i}-3\delta+1}{\sqrt{\Delta_{i}}}\arctanh\left(\frac{\sqrt{\Delta_{i}}}{\mu_{i}+\delta-1}\right)-\frac{3}{2}\log\left(\frac{\mu_{i}}{\delta}\right)\right].
\end{equation}

\section{\label{sec:fullIB}Analytic IB Amplitudes}
In this appendix, we give the analytic expressions for the three sub-amplitudes $\mathcal{A}_{\text{vb}}$, $\mathcal{A}_{\text{mix}}$ and $\mathcal{A}_{m_{f}}$ defined formally in Eq.~\eqref{eq:AIB}. A more detailed analysis can be found in~\cite{Kumar:2016cum}. In the following equations, $k_{3}$ denotes the final state fermion momentum, $k_{4}$ for the anti-fermion, and $k_{5}$ and $\epsilon_{5}$ for the photon momentum and polarization. The amplitude opened by the vector boson emission, $\mathcal{A}_{\text{vb}}$, is given by
\begin{align}
\mathcal{A}_{\text{vb}}&=\overline{u}(k_3)\,\mathcal{O}_{1}(|\lambda_{L}|^{2}\cos^{2}\alpha P_{L}-|\lambda_{R}|^{2}\sin^{2}\alpha P_{R})v(k_4)\nonumber\\*
&\quad+\overline{u}(k_3)\,\mathcal{O}_{2}(|\lambda_{L}|^{2}\sin^{2}\alpha P_{L}-|\lambda_{R}|^{2}\cos^{2}\alpha P_{R})v(k_4)\,,
\end{align}
where the matrix $\mathcal{O}_{i}$ is given by
\begin{equation}
\mathcal{O}_{i} \equiv \gamma_\mu \left[ \frac{k_{5}^\mu (k_{3}-k_{4})\cdot\epsilon_{5}- \epsilon_{5}^\mu(k_{3}-k_{4})\cdot k_{5}}{(s_{3}-m^{2}_{\widetilde{{f}_{i}}})(s_{4}-m^{2}_{\widetilde{{f}_{i}}})} \right],
\end{equation}
with $s_{3}=(k-k_{3})^{2}$ and $s_{4}=(k-k_{4})^{2}$. The mixing-induced amplitude, $\mathcal{A}_{\text{mix}}$, is given by
\begin{align}
\mathcal{A}_{\text{mix}}=m_{X}|\lambda_{L} \lambda_{R}|\sin(2\alpha)&\left[\cos\varphi\,\overline{u}(k_3)\gamma^{5}(\mathcal{V}_{1}+\mathcal{S}_{1}-\mathcal{V}_{2}-\mathcal{S}_{2})v(k_4)\right.\nonumber\\*
&\;\left.-i\sin\varphi\,\overline{u}(k_3)(\mathcal{V}_{1}+\mathcal{S}_{1}-\mathcal{V}_{2}-\mathcal{S}_{2})v(k_4)\right],
\end{align}
where the matrices $\mathcal{V}_{i}$ and $\mathcal{S}_{i}$ are
\begin{align}
\mathcal{V}_{i}
&\equiv
-\frac{i}{2} \sigma_{\mu \nu} k_5^\mu \epsilon^\nu
\left[ \frac{1}{(k_{3}\cdot k_{5})(s_{4}-m^{2}_{\widetilde{{f}_{i}}})}+\frac{1}{(k_{4}\cdot
k_{5})(s_{3}-m^{2}_{\widetilde{{f}_{i}}})} \right] \nonumber\\*
\mathcal{S}_{i}&\equiv \frac{(k_{3}-k_{4})\cdot\epsilon_{5}}{(s_{3}-m^{2}_{\widetilde{{f}_{i}}})(s_{4}-m^{2}_{\widetilde{{f}_{i}}})}+\left[\frac{k_{3}\cdot\epsilon_{5}}{(k_{3}\cdot k_{5})(s_{4}-m^{2}_{\widetilde{{f}_{i}}})}-\frac{k_{4}\cdot\epsilon_{5}}{(k_{4}\cdot k_{5})(s_{3}-m^{2}_{\widetilde{{f}_{i}}})}\right]\,.
\end{align}
Finally, the chirally suppressed piece, $\mathcal{A}_{m_{f}}$, is given by
\begin{align}
\mathcal{A}_{m_f}&=-m_{f}(|\lambda_{L}|^{2}\cos^{2}\alpha+|\lambda_{{R}}|^{2}\sin^{2}\alpha)\,\overline{u}(k_3)\gamma^{5}(\mathcal{V}_{1}+\mathcal{S}_{1})v(k_4)\nonumber\\*
&\quad-m_{f}(|\lambda_{L}|^{2}\sin^{2}\alpha+|\lambda_{{R}}|^{2}\cos^{2}\alpha)\,\overline{u}(k_3)\gamma^{5}(\mathcal{V}_{2}+\mathcal{S}_{2})v(k_4)\,.
\end{align}
If we write the momenta in the fermion pair center-of-mass frame, the differential cross section can be calculated as
\begin{equation}
    \frac{d(\sigma v)_{\text{IB}}}{dx}=\frac{x}{512\pi^{4}}\sqrt{1-\frac{m_{f}^{2}}{m_{X}^{2}(1-x)}}\int{d\Omega_{34}}\overline{\left|\mathcal{A}_{\text{IB}}\right|^{2}}\,,
\end{equation}
where the integration is over the spatial direction of the momentum $k_{3}$, which is opposite to that of $k_{4}$. The over-bar stands for summing over the final state spins while averaging over the initial state spins.

\bibliographystyle{JHEP}
\bibliography{Refs}

\end{document}